# Achieving Ultrahigh Resolution with High Efficiency: Optical Design of the 2D-RIXS Spectrometer at NanoTerasu BL02U

Jun Miyawaki

NanoTerasu Center, National Institutes for Quantum Science and Technology (QST),
468-1, Aoba, Aramaki, Aoba-ku, Sendai, Miyagi, 980-8572, Japan

## ABSTRACT

A state-of-the-art Resonant Inelastic X-ray Scattering (RIXS) facility, composed of a dedicated beamline and a 2D-RIXS spectrometer, has been constructed and commissioned at BL02U in NanoTerasu, Japan. This paper reports the optical design and optimization of this spectrometer, aiming for an ultrahigh energy resolution of <10 meV in the soft X-ray range. Conventional RIXS spectrometers using monochromatic incident X-rays suffer from a severe trade-off between energy resolution and measurement efficiency, which makes achieving resolutions of <10 meV practically unfeasible. To overcome this limitation, we adopted the $h\nu^2$ concept using dispersive incident X-rays, based on a comparative study. This approach decouples the energy resolution from the incident bandwidth. We estimate that our 2D-RIXS spectrometer improves efficiency by more than a factor of 10 compared to conventional spectrometers. The 2D-RIXS spectrometer has been successfully commissioned, demonstrating that sub-10 meV resolution measurement is achievable with practical throughput. While development continues towards the ultimate design resolution, the successful implementation validates the design strategy and offers a pathway for future high-performance RIXS instrumentation.

## I. Introduction

Resonant Inelastic X-ray Scattering (RIXS) is a photon-in/photon-out spectroscopic technique where a sample is irradiated with X-rays tuned to an inner-shell absorption edge [1,2 3,4]. The core-hole created in the absorption process is subsequently filled by a valence electron, resulting in scattered X-rays. By precisely analyzing the energy difference between the incident and scattered photons, invaluable information about low-energy excitations in the sample can be extracted. RIXS reveals a broad range of phenomena, such as charge-transfer excitations, orbital excitations (e.g., *dd*-excitations), phonons, and magnons in solid-state materials, as well as vibrations in molecular systems. RIXS is particularly powerful for investigating quantum materials with strong electron correlations, where the interplay between charge, orbital, spin, and lattice degrees of freedom determines their physical properties.

The experimental setup for RIXS comprises an X-ray source and a spectrometer to resolve the energy of the scattered X-rays. As a photon-hungry technique, RIXS requires a high-flux X-ray source, typically provided by synchrotron radiation, and various types of advanced spectrometers have been developed at facilities worldwide. Over the past two decades, remarkable advancements in energy resolution have been driven by improvements in both X-ray source performance and optical elements. This paper focuses on soft X-ray RIXS, which offers distinct advantages, such as the ability to directly explore *d*-orbital excitations via the *L* edges of 3*d* transition metals, as well as access to the *K* edges of light elements (C, N, O). These distinctive features have fueled active worldwide development efforts. As a result, the energy resolution has drastically enhanced: from approximately 1 eV in the early 2000s to around 100 meV achieved at SLS and SPring-8 in the early 2010s [5,6]. More recently, state-of-the-art facilities such as ESRF, DLS, TPS, and NSLS-II have pushed this limit further down to the 20–30 meV range [7,8,9,10].

As the scientific demand for even higher resolution grows, the inevitable decline in measurement efficiency has become the key limiting factor. In soft X-ray RIXS, this inefficiency stems from three principal factors. First, the fluorescence yield and scattering cross section in the soft X-ray range are extremely low [11], only around 0.1%. Second, the solid angle of collection for an ultrahigh-resolution spectrometer is extremely small, often less than 10 mrad in both the horizontal and vertical directions. Third, the diffraction efficiency of the grating used in such spectrometers is only a few percent. The combined probability of detecting scattered photons is therefore extremely low, on the order of $10^{-10}$, a value derived from the product of the fluorescence yield ($10^{-3}$), the acceptance angle ($\sim(10\ \text{mrad}/2\pi)^2 \approx 2.5\times 10^{-6}$), and the grating efficiency ($\sim 5\times 10^{-2}$). Even with a high incident flux of $10^{12}$ photons/s, the count rate can drop to only 100 counts/s. Pushing for <10 meV resolution worsens this problem, as narrower incident bandwidth reduces the photon flux and the grating efficiency often decreases, leading to an impractically low count rate.

Thus, achieving ultrahigh resolution requires overcoming the conflicting challenge of enhancing measurement efficiency. The optical configuration of a modern high-resolution RIXS spectrometer is fundamentally simple, typically consisting only of a grating for dispersing the energy of scattered X-rays and a two-dimensional (2D) detector for recording the intensity or counting photons of the dispersed X-rays. This simple configuration means that once a target resolution is set, the optical geometry is largely fixed, leaving few parameters available for modification to improve throughput. While various strategies to improve diffraction efficiency are being developed, such as using multilayer-coated gratings, they are beyond the scope of this paper. Considering these situations, an efficiency improvement must come from the incident beam side, specifically by using the much higher flux available in dispersive X-rays compared to highly monochromatized X-rays. Two notable spectrometer designs employ this principle: active grating monochromator – active grating spectrometer (AGM–AGS) and $hv^2$ concept [12,13]. The AGM–AGS has already demonstrated its capabilities at TLS and is operational at TPS for ultrahigh-resolution purposes [14,9]. On the other hand, the $hv^2$ concept was originally proposed by Strocov to improve measurement efficiency. This concept has been implemented at the ALS and European XFEL and planned at Hefei Advanced Light Facility, where spectrometers are designed primarily to provide imaging capabilities [15,16,17].

To meet the goal of an ultrahigh-resolution RIXS facility at NanoTerasu BL02U with a design target of <10 meV at the Cu $L$ edge, we conducted a detailed performance comparison between these two dispersive concepts. Although the AGM–AGS showed remarkable performance at specific energies, we ultimately determined to adopt the $hv^2$ concept as "2D-RIXS", prioritizing its operational versatility across a wider energy range. This paper describes the design of the 2D-RIXS spectrometer at NanoTerasu. We detail the performance comparison that guided our decision, describe the optimization of the 2D-RIXS optical design, present its expected performance, and discuss the challenges in realizing such ultrahigh resolution. The successful implementation of this instrument [18,19], which has been commissioned and is now operational, validates our design approach and provides a proven guideline for the development of future RIXS spectrometers aiming for both ultrahigh resolution and high efficiency.

## II. Numerical Evaluation of Spectrometer Energy Resolution

The energy resolution of the spectrometer was determined through a numerical analysis based on Ref. [20]. To reduce the computational cost, ray-tracing was confined to the 2D dispersion plane of the grating. A point source was employed to simulate X-rays illuminating an ideal varied-line-spacing (VLS) diffraction grating. The resulting intensity profile on the detector, calculated via ray-tracing, includes all geometrical aberrations. To evaluate the final resolution, this ideal profile was then convolved with broadening functions accounting for real-world factors, such as the finite source size, the spatial resolution of the detector, and imperfections of optical elements. The derivation of these broadening contributions and the method for calculating the total resolution are presented below.

The formula for the diffraction grating is expressed as
$$\sin\alpha + \sin\beta = a_0 k \lambda.$$
Here, $\alpha$ and $\beta$ correspond to the incident and diffraction angles relative to the grating normal, following the convention where $\alpha$ is positive and $\beta$ is negative. $a_0$, $k$, and $\lambda$ represent the groove density of the diffraction grating, the diffraction order (e.g. $k=1$ for the first inside order), and the wavelength of the X-ray, respectively.

The deviations $\Delta\alpha$ and $\Delta\beta$ give rise to corresponding change in wavelength $\Delta\lambda$, which determines the resolution. By taking partial derivatives of the equation, we can determine $\Delta\lambda$:
$$\Delta\lambda = \frac{\cos\alpha}{a_0 k}\Delta\alpha + \frac{\cos\beta}{a_0 k}\Delta\beta.$$
Since $E = \frac{hc}{\lambda}$, the energy broadening is given by $\Delta E = \frac{E^2}{hc}|\Delta\lambda|$. In the simple spectrometer configuration consisting of a source, a grating, and a detector, three key factors contribute to the broadening effect: the size of the source along the dispersion plane ($\Sigma_{\text{src}}$), the slope error of the grating ($\sigma_{\text{gr}}$), and the spatial resolution of the detector ($\sigma_{\text{det}}$), where $\Sigma$ and $\sigma$ denote root-mean-square (RMS) value. By substituting the angular deviations $\Delta\alpha = \frac{\Sigma_{\text{src}}}{r_1}$ and $\Delta\beta = \frac{\sigma_{\text{det}}}{r_2}$ (where $r_1$ and $r_2$ represent the entrance and exit arm lengths of the grating, respectively), we can derive the energy broadening contributions $\Delta E_{\text{src}}$, $\Delta E_{\text{gr}}$, and $\Delta E_{\text{det}}$ as follows:

$$\Delta E_{\text{src}} = \frac{2.64 \Sigma_{\text{src}}}{r_1} \frac{\cos \alpha}{a_0 k h c} E^2,$$

$$\Delta E_{\text{gr}} = 2.64 \sigma_{\text{gr}} \frac{\cos \alpha + \cos \beta}{a_0 k h c} E^2,$$

$$\Delta E_{\text{det}} = \frac{2.64 \sigma_{\text{det}}}{r_2} \frac{\cos \beta}{a_0 k h c} E^2.$$

The total resolution, $\Delta E_{\text{total}}$, is then calculated by convolving the ideal intensity profile from ray-tracing with a Gaussian function whose full-width at half-maximum (FWHM) is the root sum square of the three contributions: $\sqrt{\Delta E_{\text{src}}^2 + \Delta E_{\text{gr}}^2 + \Delta E_{\text{det}}^2}$. For stringent resolution evaluations, a conversion factor of 2.64 is employed to convert from RMS to FWHM [21]. Furthermore, the resolution is fundamentally limited by the diffraction limit of the grating, which is determined by the product of the total number of illuminated grooves and the diffraction order. For the present design, this limit is not a primary concern, as the grating length is sufficient to provide the required number of grooves for the target resolution and all the grooves are irradiated by the scattered X-rays in the RIXS spectrometer.

## III. Comparison of Spectrometer Optical Layout

The performance differences among RIXS spectrometers that use monochromatic X-rays have been already extensively examined [10]; thus, this study does not provide a detailed comparison of those specific designs. Instead, our investigation focuses on comparing and evaluating spectrometer concepts that employ dispersive X-rays to achieve ultrahigh resolution. As a baseline for this comparison, we first analyze a conventional spectrometer that uses monochromatic X-rays in a simple optical layout (Fig. 1(a)). We then evaluate two configurations that use dispersive X-rays: the AGM–AGS (Fig. 1(b)) and the *hv*² concept, 2D-RIXS spectrometer (Fig. 1(c)).

The primary design goal is to achieve a total resolving power of $E/\Delta E > 100{,}000$ at 1000 eV, representing the combined resolution of the beamline and spectrometer, while maintaining high measurement efficiency. Assuming a realistic design scenario, the following boundary conditions were imposed: a maximum beamline length of about 80 m, a maximum spectrometer length of about 12 m, and a requirement for a single diffraction grating to cover the 500–1000 eV energy range. The optical components are assumed to have state-of-the-art specifications, including a slope error of 0.05 µrad (RMS), a maximum grating length of 180 mm, and a detector spatial resolution of 4 µm. The incident angle to the detector is constrained to the range of 20°–40° or 140°–160°.

This section is structured as follows. First, we discuss the basic strategy for balancing resolution and throughput common to all spectrometer designs (Sec. III A). Next, we detail the specific optical designs and parameters for the conventional, AGM–AGS, and *hv*² concept models (Sec. III B). Finally, we present a quantitative comparison of their expected performance (Sec. III C).

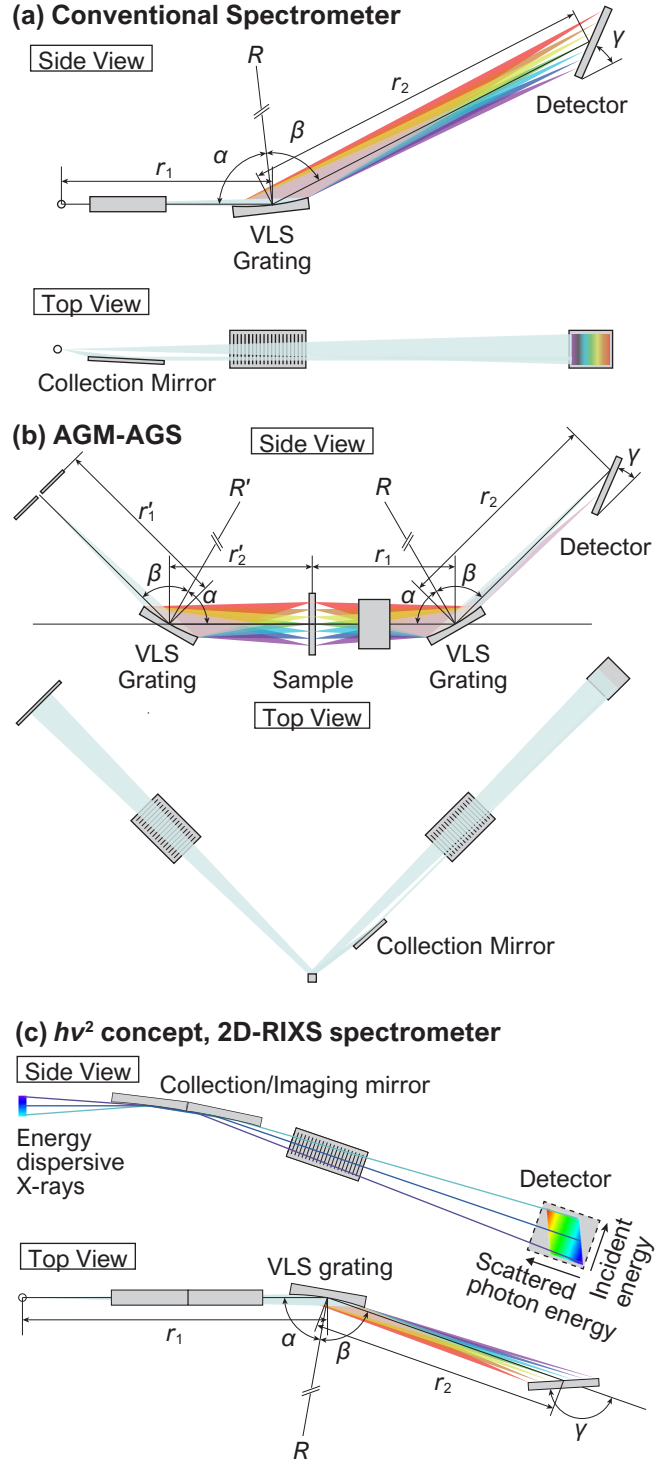

**FIG. 1.** Schematic optical layouts of the three RIXS spectrometer configurations: (a) Conventional spectrometer, (b) AGM–AGS, (c) $h\nu^2$ concept, 2D-RIXS spectrometer. Side and top views are shown for each configuration. The parameters used in the numerical optimization ($r_1$, $r_2$, $\alpha$, $\beta$, $\gamma$, and $R$) are indicated in the schematics. For AGM–AGS (b), the corresponding beamline optics are also shown, where $r'_1$, $r'_2$, and $R'$ represent the associated parameters.

## A.  Basic Strategy of Balancing Resolution and Throughput

In a conventional RIXS spectrometer, monochromatic X-rays are produced by a beamline monochromator and focused onto the sample. The scattered X-rays from the sample are then dispersed by the spectrometer grating, and the resulting spectrum is recorded by a 2D detector. The total energy resolution is a convolution of the beamline and spectrometer resolutions. To achieve a total resolving power of 100,000, each must have a resolving power greater than $100{,}000 \times \sqrt{2}$. While various combinations of resolutions can theoretically satisfy this requirement, setting equal resolving powers for both is the most practical approach to avoid imposing excessively high requirements on either the beamline or the spectrometer. The spectrometer resolution is determined by the contributions from the source size (i.e., the focus size on the sample), the grating slope error, and the detector spatial resolution. Achieving the target resolution requires a careful balance of these factors against measurement efficiency.

The contribution from the source size ($\Delta E_{\mathrm{src}}$) is inversely proportional to the entrance arm length ($r_1$). While increasing $r_1$ improves resolution, it reduces the acceptance angle of the grating and consequently lowers efficiency. For this reason, minimizing the source size (the focus size on the sample) is critical. The source size at the insertion device is usually larger horizontally than vertically, making it easier to achieve a smaller spot size in the vertical direction. For this reason, it is standard practice to orient the spectrometer dispersion plane vertically as shown in Fig. 1(a).

The grating contribution ($\Delta E_{\mathrm{gr}}$) is directly proportional to its slope error; thus, high-quality gratings are required. Furthermore, to maintain the focus widely across the detector surface for single-shot spectrum acquisition, the grating substrate must be shaped appropriately. Cylindrical gratings are commonly used for this purpose [22].

The detector contribution ($\Delta E_{\mathrm{det}}$) improves as its spatial resolution becomes smaller. However, with the best spatial resolutions of a few micrometers, even with post-processing techniques like centroiding [23], achieving the target energy resolution is challenging. The resolution can be improved by increasing the exit arm length ($r_2$), which magnifies the dispersion to the detector.

Finally, to improve measurement efficiency, the limited acceptance angle of the spectrometer must be addressed. It is now common practice to install collection mirrors between the sample and the spectrometer grating. This mirror, oriented orthogonally to the dispersion plane of the spectrometer, collects X-rays from a much larger solid angle in the non-dispersive direction [24]. This collection can be achieved without, in principle, degrading the energy resolution. This approach provides crucial efficiency gains, and by using a parabolic mirror, $r_1$ and $r_2$ can be flexibly adjusted to optimize the resolution.

## B. Spectrometer Design

### 1. Conventional Spectrometer

We begin by establishing the optical parameters for a conventional spectrometer designed to achieve a total resolving power of $E/\Delta E > 100{,}000$. To create a relevant baseline for direct comparison with the dispersive concepts, this conventional model uses the same spectroscopic optical configuration as the 2D-RIXS spectrometer (Sec. III B 3), but with the entire spectrometer rotated by 90° around the optical axis. The resulting parameters for the spectrometer are shown in Table I.

Although this configuration, which uses a plane grating ($R=\infty$) and a large detector angle ($\gamma=160°$), is somewhat unique for a conventional design, it is analogous to the spectrometer type referred to as SLVS-B in Ref. [10]. While the SLVS-B is defined as a spectrometer using a spherical grating with variable entrance and exit arm lengths ($r_1$ and $r_2$), our design corresponds to the limiting case where the radius of curvature is infinite ($R=\infty$). Further details on these parameters are available in the dedicated section on the 2D-RIXS design (Sec. IV B). The collection mirror was designed following the optimization procedure described in Ref. [8], which uses the available widths of the diffraction grating and detector as key parameters. This analysis led to the selection of a single long parabolic mirror over two shorter ones to maximize collection efficiency.

TABLE I. Optical parameters of the three RIXS spectrometer configurations at 1000 eV: Conventional, AGM–AGS, and 2D-RIXS. The table summarizes the parameters for the source, grating, detector, and collection mirror. For the AGM–AGS, the parameters for the corresponding beamline are also included.z

|  | Parameters | Symbol and units | Conventional | AGM–AGS | 2D-RIXS |
|---|---|---|---|---|---|
| **Spectrometer** | | | | | |
| Source | Focus size at sample along grating dispersion direction | $\Sigma_{src}$ (μm in FWHM) | <1 (vertical) | <0.6 (vertical) | <1 (horizontal) |
| Grating | Entrance arm length | $r_1$ (mm) | 3441 | 2500 | 3441 |
|  | Exit arm length | $r_2$ (mm) | 8559 | 9500 | 8559 |
|  | Incidence angle | $\alpha$ (°) | 89.12 | 85.84 | 89.12 |
|  | Diffraction angle | $\beta$ (°) | 84.98 | 89.00 | 84.98 |
|  | Shape of substrate |  | Plane | Cylindrical | Plane |
|  | Meridional radius of curvature | $R$ (mm) | — | 87651.2 | — |
|  | slope error | $\sigma_{gr}$ (μrad in RMS) | 0.05 | 0.05 | 0.05 |
|  | diffraction order | $k$ | 3 | −1 | 3 |
|  | Groove density | $a_0$ (lines/mm) | 1000 | 2000 | 1000 |
|  | VLS parameters | $a_1$ (lines/mm$^2$) | 0.2590 | −0.8941 | 0.2590 |
|  |  | $a_2$ (lines/mm$^3$) | 3.4024×10$^{-5}$ | 6.3803×10$^{-4}$ | 3.4024×10$^{-5}$ |
|  |  | $a_3$ (lines/mm$^4$) | 9.5739×10$^{-9}$ | −3.5125×10$^{-7}$ | 9.5739×10$^{-9}$ |
| Detector | Incidence angle | $\gamma$ (°) | 155.1 | 25 | 155.1 |
|  | spatial resolution | $\sigma_{det}$ (μm in FWHM) | 4 | ← | ← |
| Collection/ Imaging mirror | Shape |  | Parabola | Parabola | Wolter type I (Hyperbola, ellipse) |
|  | Distance from source to mirror | (mm) | 192 | ← | 150.72, 520.85 |
|  | Distance from mirror to focus | (mm) | ∞ | ← | 290.43, 11618.9 |
|  | Eccentricity | $e$ | 1 | 1 | 1.001593, 0.9999630 |
|  | Distance from focus to directrix | $p$ (mm) | 0.75098 | ← | 0.22235, 0.44877 |
|  | Semi-major axis/ semi-minor axis | $a/b$ (mm) | — | — | $a$=69.856/$b$=3.944, $a$=6069.86/$b$=52.191 |
|  | Incident angle | (°) | 88.73 | ← | 88.92, 88.78 |
|  | Working distance | (mm) | 100 | ← | 100 |
|  | Length | (mm) | 400 | ← | 400 (228+172) |
|  | Parabola, ellipse, and hyperbola are expressed by $r = \frac{ep}{1-e\cos(\theta)}$. | | | | |
| **Beamline** | | | | | |
| Grating | Entrance arm length | $r_1'$ (mm) | — | 4 | — |
|  | Exit arm length | $r_2'$ (mm) | — | 2.5 | — |

## 2. AGM–AGS

The AGM–AGS concept, employed at TLS and TPS [12,14,9], achieves high resolution by using diffraction orders of opposite signs to create a symmetric optical layout for the beamline and spectrometer, as depicted in Fig. 1(b). This symmetry results in the precise compensation of the energy dispersion from the beamline monochromator by the inverse dispersion in the spectrometer. For elastic scattering, X-rays of different energies from the beamline that are focused onto different positions on the sample are all refocused to the exact same position on the detector of the spectrometer.

This energy compensation mechanism breaks the standard convolutional relationship between the beamline and spectrometer resolutions. The energy dispersion introduced by the beamline monochromator is cancelled by the spectrometer. As a result, the energy resolution becomes independent of the bandwidth defined by the exit slit, effectively removing the contribution of the exit slit size from the equation. The total energy resolution is instead determined by the focusing performance of the beamline and spectrometer gratings. Specifically, it is limited by two main factors: (1) the focus size on the sample, which is determined by the beamline focusing capability and beamline grating slope error, and (2) the optical imperfections of the spectrometer, namely its grating slope error and finite detector spatial resolution. In other words, the total resolution is determined by the spectrometer parameters alone, since the focus on the sample acts as the effective source size for the spectrometer.

A vital design choice in the AGM–AGS is the selection of diffraction orders. The optimal combination in the AGM–AGS would be a positive diffraction order in the beamline monochromator and a negative order in the spectrometer. Using a negative order in the spectrometer is advantageous as it increases the grating acceptance angle and magnifies the dispersion onto the detector, relaxing the requirement for the spatial resolution of the detector. However, this choice increases the resolution sensitivity to the source size, demanding either a smaller X-ray focus on the sample or a longer spectrometer entrance arm ($r_1$). The reverse combination (negative order in the beamline, positive in the spectrometer) is impractical for two main reasons. First, it severely limits the achievable resolution, as it makes beamline focusing difficult. Second, it drastically reduces total throughput due to the smaller grating acceptance of both the beamline and spectrometer. For the beamline monochromator, achieving ultrahigh resolution with a negative diffraction order requires a highly demagnifying geometry ($r_1 \gg r_2$) with a maximally grazing incidence angle. Inevitably, the inherent advantage of a large acceptance angle usually associated with negative orders is effectively lost.

The use of a negative diffraction order in the spectrometer introduces considerable challenges. First, controlling the focal plane becomes difficult. To detect scattered X-rays at the detector, the incidence angle to the detector must be larger than approximately 20°; however, a plane grating in this configuration produces a focal plane (which corresponds to the incident angle to the detector) tilted at less than 1°. This necessitates the use of a cylindrical grating to adjust the focal plane. Although a cylindrical substrate makes the focal plane upright at a specific energy, the grazing exit geometry ($\beta \approx -90°$), which is required for an upright focal plane with a negative diffraction order, causes the focal distance to be highly sensitive to energy changes. This means that the energy range over which the focus is maintained is extremely narrow. It is therefore challenging to cover a broad energy range (e.g., 500–1000 eV) with a fixed-curvature grating. Achieving ultrahigh resolution over a wide incident energy range requires a grating with variable curvature, a major technical challenge [9].

The most fundamental limitation of the AGM–AGS concept for achieving ultrahigh resolution lies in its symmetric design constraint. To improve the spectrometer resolution, one must increase its entrance arm length ($r_1$). However, due to the symmetric layout ($r_2' = r_1$), as shown in Fig. 1(b), this simultaneously increases the beamline exit arm length ($r_2'$), which in turn enlarges the X-ray focus size on the sample. This enlargement degrades the resolution, counteracting the benefit of a longer $r_1$. Therefore, ultrahigh resolution can only be pursued by shortening both $r_1$ and $r_2'$ to achieve the smallest possible focus on the sample. Accordingly, the layout of an ultrahigh resolution AGM–AGS tends to resemble that of the RIXS beamline at TPS [9]. Following this approach, we adopt parameters of $r_1'$=4 m, $r_2'$=$r_1$=2.5 m, and $r_2$=9.5 m. To achieve a resolving power of >100,000, a high groove density of $a_0 \geq 2000$ lines/mm is required. The resulting optical parameters for the AGM–AGS model are summarized in Table I. The parameters for the collection mirror are the same as that used for the conventional spectrometer.

With these parameters, the usable range of dispersed X-ray energies is examined. First, let us consider the energy dispersion from the beamline at the sample position. Since the beamline grating uses a positive diffraction order and the focal plane is set to align with the sample surface (vertical plane) at the optimized energy, the focal plane remains upright over a broad energy range 1000±5 eV (spatially ±0.43 mm), as shown in Fig. 2(a). Thus, the energy dispersion of the beamline is not a limiting factor.

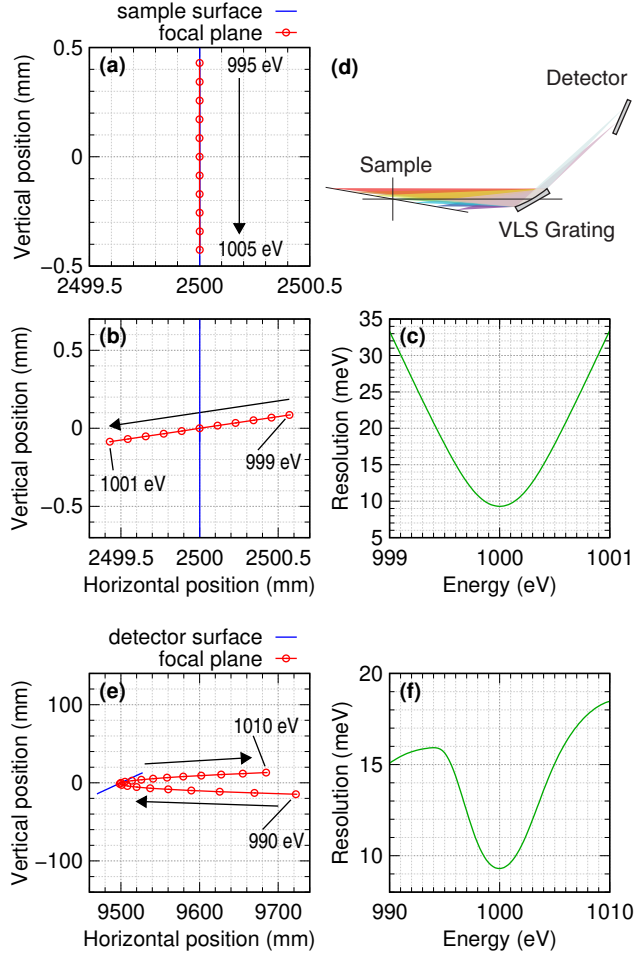

**FIG. 2.** Focal planes and energy resolution dependence for the AGM–AGS configuration. (a) Focal plane of the beamline monochromator at the sample position. (b) Reverse focal plane of the spectrometer at the sample position obtained by reverse ray-tracing. (c) Energy resolution dependence on the sample derived from the reverse focal plane analysis. (d) Schematic illustration of the mismatch between the vertical sample surface and the tilted reverse focal plane of the spectrometer. (e) Focal plane on the detector calculated by forward ray-tracing. (f) Energy resolution dependence on the detector.

On the other hand, the usable range is limited by how well the spectrometer can maintain focus for a spatially extended and energy-dispersed source, i.e., the off-axis focusing capability of the spectrometer grating. To understand this condition, it is useful to consider the "reverse focal plane" at the sample position based on the principle of the reversibility of light. Let us first confirm the parameters for the standard forward direction from the sample through the grating to the detector. The focal plane at the detector is set to 25°, and the grating curvature and VLS parameters were determined to make the focal plane match this plane. As shown in Table I, the parameters are $r_1$=2.5 , $r_2$=9.5 m, $k$=−1, with $R$=87651.1 mm, $a_0$=2000 lines/mm, $a_1$=−0.8941 lines/mm$^2$, $a_2$=6.3803×10$^{-4}$ lines/mm$^3$, and $a_3$=−3.5125×10$^{-7}$ lines/mm$^4$. For a source size of 0.6 μm, the purely geometrical focus size at the detector position is 9.47 μm, assuming an ideal grating surface and measured on a plane normal to the optical axis. To evaluate the off-axis focusing capability on the source side, we can simply consider these parameters in the reverse direction: $r_1$=9.5 m, $r_2$=2.5 m, $k$=+1, with the same $R$ and $a_0$, but with the signs of $a_1$ and $a_3$ inverted ($a_1$=+0.8941 lines/mm$^2$, $a_3$=+3.5125×10$^{-7}$ lines/mm$^4$). Note that $r_1$ and $r_2$ are swapped, and the signs of $k$, $a_1$, and $a_3$ are reversed. By ray-tracing with a source size of 9.47 μm in this reverse configuration, we obtained the reverse focal plane for the range of 999.5–1000.5 eV, as shown in Fig. 2(b). The focal plane is severely tilted at 8.5° relative to the optical axis, whereas the sample surface is oriented at 90° (vertical). As a result, while the resolution at 1000 eV achieves the design value of 9.3 meV, it degrades to 10.8 meV at 1000±0.2 eV (±25 μm) (Fig. 2(c)). This situation is schematically illustrated in Fig. 2(d); the grating in this setup can only successfully focus scattering that originates from a plane tilted at 8.5°, meaning it cannot effectively focus scattering from the ideal vertical plane shown in Fig. 1(b). In AGM–AGS, scattered

X-rays of different energies are focused into the same position on the detector, resulting in an averaged and blurred focus, but it is evident that using a wide energy range with the ultrahigh resolution is difficult. This behavior has been confirmed not only by the reverse evaluation but also by full ray-tracing simulations using SHADOW [25].

Furthermore, Fig. 2(e) shows the focal plane on the detector for the range of 990–1010 eV using the parameters in Table I. Although the range of approximately −33 to +30 mm may exceed the detector size, the figure clearly demonstrates the difficulty of flattening the focal plane with a negative diffraction order. The corresponding resolution at each energy is shown in Fig. 2(f). The energy resolution is 9.3 meV at the optimum energy of 1000 eV, but degrades to 10.6 meV at ±2 eV and further to 15.7 meV at ±5 eV. This directly impacts the available energy loss range, indicating that the region where <10 meV resolution is limited to a bandwidth of approximately 4 eV. In the negative diffraction order geometry, it is particularly difficult to flatten the focal plane across a broad energy range.

It would be desirable to resolve this difficulty through parameter optimization, but it is fundamentally constrained. This limitation arises because the system is over-constrained: three conditions (the off-axis focusing capability at the sample, the focus and focal plane angle at the detector) must be optimized using only two independent parameters: the grating curvature $R$ and the VLS parameter $a_1$. Therefore, satisfying all three conditions simultaneously is mathematically over-constrained. Although it is possible to make the reverse focal plane somewhat more upright by adopting a maximally grazing incidence angle for the grating, satisfying all conflicting requirements simultaneously remains difficult with a negative order configuration.

### 3. $h\nu^2$ concept — 2D-RIXS spectrometer

The $h\nu^2$ concept, originally proposed by Strocov, arranges the dispersion directions of the beamline monochromator and the spectrometer to be orthogonal, as depicted in Fig. 1(c). The spectrometer features two key functions: a 1D imaging capability to resolve the spatial distribution of the energy-dispersed incident X-rays, and a conventional spectroscopic capability to resolve the energies of the scattered X-rays in the orthogonal direction. This configuration enables simultaneous 2D measurement of both incident and scattered photon energies. Based on this capability, we have named the instrument at NanoTerasu BL02U the "2D-RIXS" spectrometer. Utilizing this imaging capability, the $h\nu^2$ concept extends its utility beyond the standard RIXS spectrometer, enabling single-shot energy-dependent measurements like dispersive XAFS or spatial-dependence measurements akin to microscopy. By integrating the RIXS spectra recorded over a narrow range of incident energies, the measurement efficiency can be greatly enhanced, similar in principle to the AGM–AGS concept [26].

The primary requirement of this layout is the orthogonality of the dispersion directions in the beamline and spectrometer. While the combination of a horizontally dispersive beamline and a vertically dispersive spectrometer is possible, achieving ultrahigh resolution practically requires a vertically dispersive beamline monochromator to take advantage of the smaller vertical source size of the synchrotron. This necessitates a horizontally dispersive spectrometer.

The 2D-RIXS spectrometer in turn imposes several requirements on the beamline: a resolving power of $E/\Delta E \geq 100{,}000 \times \sqrt{2}$; a well-defined vertical focal plane that is aligned with the sample surface and a horizontal focus size below 1 μm, which defines the effective source size for the spectrometer. For the focal plane requirement, a standard plane grating monochromator (PGM) is sufficient given that a vertical line of only 600 μm (at maximum) of the dispersed beam is used. This will be explained in detail in the next section (Sec. IV A). Achieving a stable horizontal focus below 1 μm is challenging but feasible with optimized beamline design using Wolter type I mirror for the refocusing [27,18].

In the 2D-RIXS spectrometer, the one-dimensional (1D) imaging function should be realized using achromatic imaging optics, such as Wolter mirrors. The achromatic nature of the Wolter mirror imposes an additional constraint on the spectrometer design: to maintain focus for all X-ray energies, the total path length from the sample to the detector must be held constant ($r_1+r_2=12$ m). The spectroscopic performance of the 2D-RIXS spectrometer was optimized following the methodology described in Ref. [20] (Sec. IV B). Details on the optimization of the Wolter mirrors are provided in a later section (Sec. IV C). The resulting optical parameters for the 2D-RIXS spectrometer, which represent the actual parameters of the spectrometer at NanoTerasu BL02U, are summarized in Table I.

### C. Comparison of Performance

Based on these designs detailed above, a quantitative comparison of throughput at 1000 eV for the three spectrometer concepts is summarized in Table II, with performance data from the NSLS-II SIX beamline included for the comparison [10]. The total throughput was calculated from the beamline flux, the effective acceptance angles of the collection mirror and grating (defined as the physical acceptance angle including the effect of reflectivity), mirror reflectivity (for SIX only), grating diffraction efficiency, and the usable range of the beamline energy dispersion, referred to as the effective field of view (eFOV), which is defined as the range maintaining the ultrahigh energy resolution. Performance data for SIX were taken directly from Ref. [10]. It should be noted that all values presented in this comparison are calculated based on simulations, rather than experimental measurements. The beamline flux for the conventional, AGM–AGS, and 2D-RIXS models was calculated assuming the NanoTerasu BL02U configuration (two focusing mirrors, monochromator (plane mirror + grating), refocusing mirror(s)). As expected, the conventional RIXS model suffers from impractically low throughput, calculated to be about one-tenth of that achieved by the SIX ultrahigh-resolution mode. In contrast, both dispersive concepts offer substantial gains. The AGM–AGS design is calculated to have a throughput over 50 times higher than SIX, while the 2D-RIXS design is slightly better than SIX. This demonstrates the clear advantage of using dispersive X-rays to improve measurement efficiency.

The high throughput of the AGM–AGS concept, however, comes at the cost of considerable operational complexity and limitations. The primary challenge is the necessary use of a negative diffraction order in the spectrometer for the symmetric configuration. This choice complicates focal plane control and severely restricts the usable energy range for a fixed-curvature grating. For example, the focal plane tilt at the detector can become close to 0° when changing the energy from the reference energy, and the usable energy range of dispersed X-rays on the sample is limited to a few hundred meV. Overcoming these issues requires a grating with variable curvature, a major technical challenge, especially when considering the stringent slope error tolerances required for ultrahigh resolution [9].

The $h\nu^2$ concept circumvents these issues by separating the imaging and spectroscopic functions into orthogonal planes. The 1D imaging is performed by an achromatic Wolter mirror, whose alignment is energy-independent and has relatively large tolerances. Once optimized, the adjustment would not often be required, provided mechanical stability is maintained. The spectroscopic function, although rotated by 90°, is fundamentally identical to that of a conventional spectrometer, relying on well-understood principles. This separation provides distinct operational advantages.

**TABLE II.** Quantitative comparison of the calculated throughput at 1000 eV. Data for SIX were taken from Ref. [10]. Grating efficiency is calculated by the differential method [28].

| | | BL grating (lines/mm, order) | Exit slit width (μm) | Res. of BL (meV) | Photon flux at sample (ph/s) | Coll. mirror acc. (mrad) | Grating acc. (mrad) | VFM refl. | PM refl. | Spectr. grating (lines/mm, order, type) | Grating efficiency | Net res. (meV) | eFOV (μm) | gain by dispersion | Relative Through-put |
|---|---|---|---|---|---|---|---|---|---|---|---|---|---|---|---|
| SIX | Medium | 500, +1 | | 60 | 9.0×10$^{12}$ | 36 | 5 | 0.7 | 0.66 | 1250, +1, blazed | 0.15 | 86 | — | — | 183.3 |
| | High | 1200, +1 | | 23.5 | 2.0×10$^{12}$ | 36 | 5 | 0.7 | 0.54 | 2500, +1, blazed | 0.06 | 31 | — | — | 13.3 |
| | Ultrahigh | 1800, +1 | | 10 | 1.5×10$^{11}$ | 36 | 5 | 0.7 | 0.54 | 2500, +1, blazed | 0.06 | 14 | — | — | 1 |
| Conventional | | 900, +2 | 2 | 6.5 | 3.8×10$^{10}$ | 49 | 0.802 | — | — | 1000, +3, blazed | 0.041 | 10 | — | — | 0.101 |
| AGM–AGS | | 2000, +1 | 0.79 | 10 | 1.2×10$^{11}$ | 49 | 4.33 | — | — | 2000, +1, laminar | 0.048 | 10 | 30 | 38 | 73.4 |
| 2D-RIXS | High | 1000, +1 | 2 | 16.8 | 1.2×10$^{11}$ | 10 | 1.26 | — | — | 1400, +1, laminar | 0.087 | 22.7 | 160 | 80 | 17.1 |
| | Ultrahigh | 900, +2 | 2 | 6.5 | 4.1×10$^{10}$ | 10 | 0.802 | — | — | 1000, +3, blazed | 0.041 | 10 | 160 | 80 | 1.78 |

The $h\nu^2$ concept also offers unique measurement capabilities. Because the incident energy is resolved spatially on the detector, it is possible to measure the incident energy dependence of the RIXS spectra in a single acquisition. If this dependence is negligible, the data can be integrated to improve statistics, or alternatively, used to perform spatially-resolved RIXS microspectroscopy, "RIXS microscopy" [15,16,29]. In contrast, the AGM–AGS design, where scattered X-rays with the same energy loss are focused to the same position on the detector, cannot provide this incident-energy or spatial information.

Although AGM–AGS offers high throughput, it requires a variable curvature grating to achieve ultrahigh resolution over a wide energy range, which introduces operational complexities and alignment challenges. In contrast, the $h\nu^2$ concept provides greater operational simplicity and unique measurement capabilities. Therefore, the $h\nu^2$ concept was ultimately selected for NanoTerasu BL02U. While the AGM–AGS concept remains a sophisticated and powerful technique for achieving maximum throughput at a fixed incident energy, the 2D-RIXS approach was deemed to offer a more robust and flexible strategy for realizing ultrahigh resolution RIXS in practical experiments across diverse absorption edges.

## IV. Design of 2D-RIXS Spectrometer at NanoTerasu BL02U

This section details the design of the 2D-RIXS spectrometer constructed at NanoTerasu BL02U. A schematic diagram of the instrument is shown in Fig. 1(c). Based on the $h\nu^2$ concept, vertically energy-dispersed X-rays from the beamline monochromator are incident on the sample. The spectrometer then performs two separate functions simultaneously on orthogonal planes. In the vertical plane, an imaging optics magnifies and images the scattered X-rays onto the detector, preserving their spatial relationships. In the horizontal plane, a diffraction grating disperses the scattered X-rays by energy and focuses them onto the detector. While the spectroscopic function is rotated by 90° around the optical axis relative to a conventional spectrometer, its fundamental principles are identical. The orthogonality of the imaging and spectroscopic optics means that they do not interfere, allowing for their independent design, evaluation, and optimization.

The following subsections will detail this design. First, we will briefly introduce the beamline that serves as the source for the 2D-RIXS spectrometer (Sec. IV A). We will then describe the design and optimization of the spectroscopic optics in the horizontal plane (Sec. IV B) and the imaging optics in the vertical plane (Sec. IV C). Finally, we will evaluate the expected total performance of the complete 2D-RIXS spectrometer through full ray-tracing simulations (Sec. IV D) and discuss the mechanical stability required to achieve ultrahigh resolution (Sec. IV E).

### A. Beamline

NanoTerasu BL02U is designed for the ultrahigh-resolution 2D-RIXS spectrometer, aiming for a combined resolving power of $E/\Delta E > 100{,}000$ through the integrated optical design of the beamline and spectrometer. The detailed optical design of the beamline is described elsewhere [27]; here, we summarize the key features relevant to the spectrometer.

Figure 3 shows the schematic optical layout of BL02U. The beamline employs a focusing varied-line-spacing plane grating monochromator (FVLSPGM) [30] composed of M2 and BL-G and is equipped with entrance slits (S1V and S1H) but no exit slit. Upstream of the monochromator, two mirrors (M0 and M1) initially focus the beam: the first mirror (M0) focuses the X-rays from the undulator vertically to the vertical entrance slit (S1V), while the second mirror (M1) focuses them horizontally to the horizontal entrance slit (S1H). In the vertical direction, the focus at S1V acts as the effective source for the FVLSPGM. The grating (BL-G) then disperses the X-rays by energy and focuses them directly onto the sample, which is located at the nominal exit slit position. In the orthogonal horizontal direction, a Wolter type I mirror (M3) serves as the final refocusing optic. It demagnifies and images the focus at S1H onto the sample, delivering a horizontal focus of <1 μm. The footprint of this spot on the sample defines the effective source for the spectroscopic optics of the spectrometer. As a result, the beam profile on the sample is a vertically elongated line due to the combination of vertical energy dispersion and tight horizontal focusing. The designed and measured parameters of the four mirrors (M0, M1, M2, and M3) and the beamline gratings are summarized in Table III. Two types of gratings are installed: the BL-G900 (900 lines/mm, used in the 2nd order) for ultrahigh resolution and the BL-G1000 (1000 lines/mm, used in the 1st order) for high resolution.

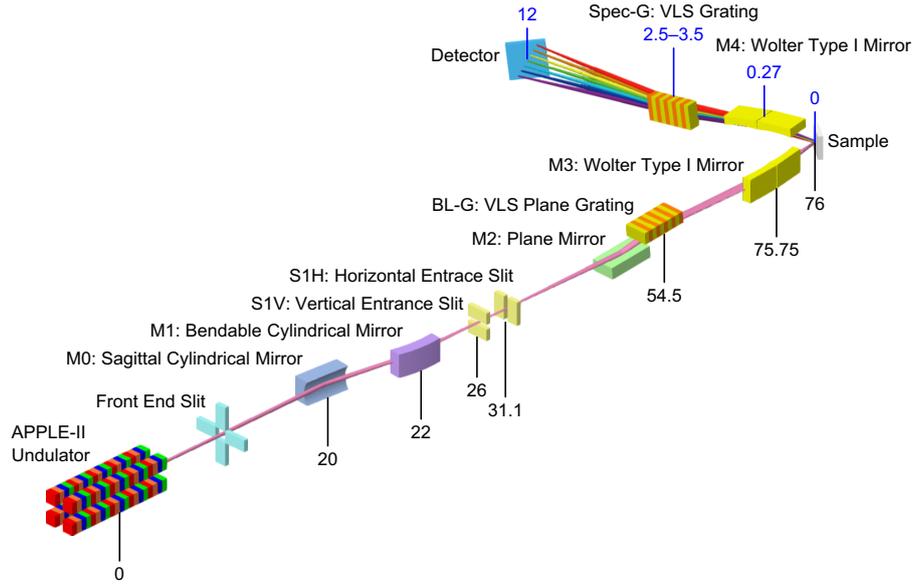

**FIG. 3.** Schematic optical layout of the NanoTerasu BL02U and 2D-RIXS spectrometer. The numbers below and above the optical elements indicate the distance from the source (The undulator for the beamline and the sample for the spectrometer).

**TABLE III.** Mirror and grating parameters of the beamline. Values in brackets [ ] are the measured data.

|  | M0 | M1 | M2 | M3 | |
|---|---|---|---|---|---|
|  |  |  |  | Wolter type I | |
| Shape | Sagittal Cylinder | Bendable cylinder | Plane | Ellipse | Hyperbola |
| Entrance arm length (m) | 20 | 22 | — | 44.532 | 0.3178 |
| Exit arm length (m) | 6 | 9.1 | — | 0.5389 | 0.1471 |
| Angle of incidence (°) | 88.8 | 88.8 | 85.1–88.9 | 88.74 | 88.56 |
| Meridional radius | >50 km [∞] | Variable (to be used at 614.76 m) | >140 km [1767.0 km] | $a$=22535.33 mm $b$=107.40 mm | $a$=85.326 mm $b$=5.445 mm |
| Sagittal radius | 193.331 mm [193.375 mm] | >1500 m [∞] | >1500 m [∞] | — | — |
| Material of substrate | Si | Si | Si | Si | |
| Meridional slope error (μrad in RMS) | <0.2 [0.172] | <0.1 [0.097] | <0.05 [0.036] | <0.15 [0.085] | <0.15 [0.079] |
| Sagittal slope error (μrad in RMS) | <5.0 [0.364] | <1.0 [0.352] | <1.0 [0.022] | <1.0 [0.089] | <1.0 [0.155] |
| Material of coating | Au | Au, bare Si | Au | Au | Au |
| Useful area $L \times W$ (mm²) | 180×10 | 180×10 ×2 stripes | 380×20 | ~239×20 | ~139×20 |
| Manufacturer | JTEC | JTEC | JTEC | JTEC | JTEC |

|  | BL-G900 | BL-G1000 |
| --- | --- | --- |
| Energy range (eV) | 500–1000 | 250–2000 |
| Shape of substrate | Plane | Plane |
| Reference energy (eV) | 1000 | 1000 |
| Incident angle at reference energy (°) | 89.1 | 89.0 |
| Meridional radius (m) | >140 km [666.4 km] | >140 km [∞] |
| Sagittal radius (mm) | >20 km [∞] | >20 km [∞] |
| Meridional slope error (μrad in RMS) | <0.05 [0.045] | <0.1 [0.072] |
| Sagittal slope error (μrad in RMS) | <0.5 [0.079] | <0.5 [0.059] |
| Groove profile | Blazed | Laminar |
| Diffraction order | 2 | 1 |
| Ruled area $L \times W$ (mm$^2$) | 180×20 | 180×14×2 stripes |
| Blaze angle (°) | 1.6 [1.59] | — |
| Duty ratio | — | 0.3 [0.34] |
| Groove depth (nm) | — | 10 [12.3], 16 [17.8] |
| Material of substrate | Si | Si |
| Material of coating | Au | Au |
| VLS parameters: |  |  |
| $a_0$ (lines/mm) | 900 | 1000 [1000.063] |
| $a_1$ (lines/mm$^2$) | 0.0917 | 0.113 [0.1128] |
| $a_2$ (lines/mm$^3$) | 5.96×10$^{-6}$ | 6.82×10$^{-6}$ [7.221×10$^{-6}$] |
| $a_3$ (lines/mm$^4$) | 3.88×10$^{-10}$ | 4.71×10$^{-10}$ |
| Manufacturer | JTEC/Inprentus | JTEC/Shimadzu |

Here, we discuss in detail two critical design choices of the beamline that are essential for the performance of the 2D-RIXS spectrometer. The first concerns the focal plane of the beamline grating. Since the sample is mounted vertically, it is ideal for the focal plane to be vertical (upright, 90°) to align with the sample surface. However, the focal plane of the FVLSPGM adopted for BL02U is tilted with respect to the optical axis. Fig. 4(a) shows the energy dependence of the focal plane angle for the two gratings, where 180° corresponds to the horizontal plane. In the ultrahigh resolution mode using the BL-G900 grating, the focal plane is relatively upright, with angles ranging from 166° to 147°. In contrast, in the high resolution mode using the BL-G1000 grating, the focal plane is steeply inclined towards the horizontal, with angles ranging from only 175° to 166°. Nevertheless, this does not compromise the performance because the effective FOV of the spectrometer, defined as the range where a spatial resolution of ≤2 μm is maintained to achieve the ultrahigh energy resolution, is limited to approximately 200–300 μm (whereas the full range determined by the detector size is approximately 600 μm). Figure 4(b) and (c) plot the focus size and energy resolution dependence calculated around 1000 eV using BL-G900 and BL-G1000. It is evident that the FVLSPGM configuration does not compromise the energy resolution; the defocus caused by the focal plane tilt is negligible within this narrow range. This is due to the long exit arm of the monochromator (21.5 m); even with a tilt of, for example, 10°, a vertical deviation of ±0.3 mm corresponds to a longitudinal shift of only about 1.7 mm, which falls well within the focal depth. Figure 4(d) shows the energy dispersion at the sample position as a function of photon energy. Based on these considerations, we prioritized the FVLSPGM design to maximize the effective use of the beamline length.

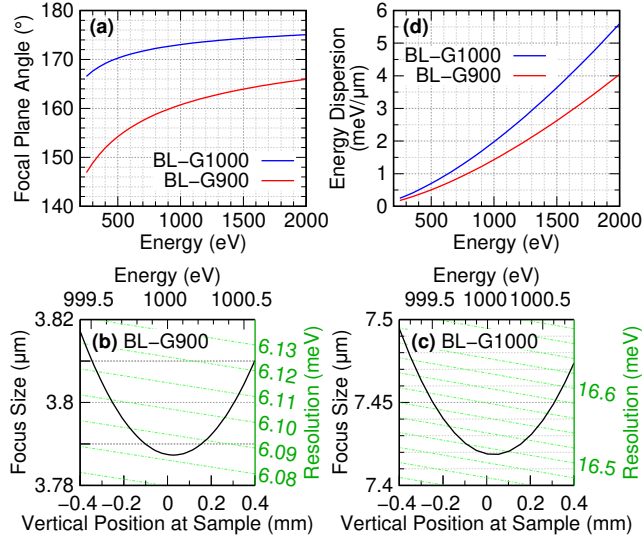

**FIG. 4.** Calculated characteristics of the beamline optics. (a) Energy dependence of the focal plane angle for BL-G900 and BL-G1000. 180° corresponds to the horizontal plane. (b), (c) Vertical position dependence of the focus size around 1000 eV for BL-G900 and BL-G1000, respectively. The focus size is also expressed in the energy scale. (d) Energy dependence of the energy dispersion.

The second key point is the selection of a Wolter type I mirror for the final horizontal refocusing. Since RIXS is a photon-hungry technique, maximizing flux is generally a priority, which would favor a single-reflection elliptical mirror over a double-reflection Wolter mirror. However, for BL02U, we prioritized focusing quality, reliability, and stability over throughput. Because the horizontal source size of synchrotron radiation is considerably larger than the vertical size, achieving a focus of <1 μm requires an exceptionally high demagnification ratio. At BL02U, this is achieved in two stages (M1 and M3), with a total demagnification of 432.2. The final refocusing mirror (M3) alone must provide a demagnification of 178.6. At such a high demagnification, a single elliptical mirror would suffer from aberrations, degrading the focus quality (Fig. 5(a)). In contrast, the Wolter mirror can achieve this tight focus with practically zero aberration (Fig. 5(a)). Furthermore, the Wolter mirror offers an angular tolerance approximately 50 times larger than that of an elliptical mirror (Fig. 5(b)), providing distinct advantages in terms of ease of alignment and long-term stability. Since the horizontal focus size and shape directly determine the energy resolution of the spectrometer, establishing a clean and stable source is crucial. As demonstrated in Ref. [18], the horizontal focusing is uniform across a vertical range of ±300 μm, consistent with simulations. Therefore, the Wolter mirror stably delivers the expected vertically elongated line profile with a highly symmetric horizontal focus of <1 μm, which is one of the key factors in achieving the ultrahigh resolution.

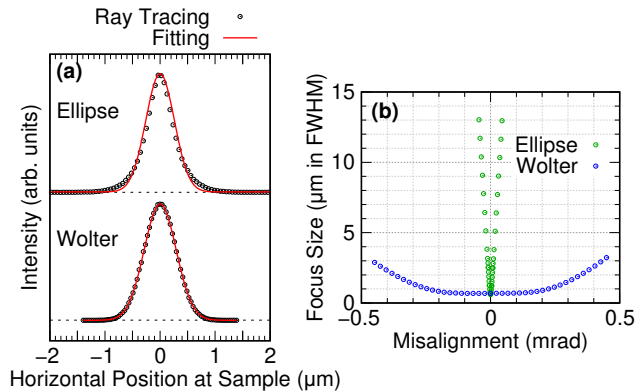

**FIG. 5.** Comparison of focusing performance between a Wolter type I mirror and a single elliptical mirror for M3. (a) Simulated focusing profiles. Open circles and red lines represent ray-tracing data and fitting results, respectively. (b) Comparison of tolerance for misalignment in the incident angle.

## B. Spectroscopic Optics in 2D-RIXS Spectrometer

The boundary conditions for the design of the spectroscopic optics are summarized in Table IV. The effective source size for the spectrometer is determined by the horizontal focus size of the beamline, which is set to be less than 1 μm. The state-of-the-art slope error for the optics is assumed to be 0.05 μrad in RMS. The spatial resolution of the detector, using centroiding techniques, is set to 4 μm at 1000 eV. As previously mentioned, the vertical imaging optics require a constant total optical path length due to their achromatic nature; hence, the spectroscopic optics must also maintain a constant path length. Due to space constraints at BL02U, this total length ($r_1+r_2$) is set to 12 m. The mechanical range of motion for the diffraction grating ($r_1$) is 2500–3500 mm (a range of approximately 1 m). The included angle of the grating ($\alpha-\beta$) is constrained between 176° and 168° (a variation of 12°), and the incident angle on the detector surface ($\gamma$) is allowed to vary by 20°.

TABLE IV. Boundary conditions and constraints for the design of the spectroscopic optics in the 2D-RIXS spectrometer.

| | |
|---|---|
| Source size (μm) | 1 |
| Slope error of grating (μrad in RMS) | 0.05–0.10 |
| Spatial resolution of detector (μm) | 4 |
| Entrance arm length of grating, $r_1$ (mm) | 2500–3500 |
| Optical path length, $r_1+r_2$ (mm) | 12000 |
| Included angle of grating, $\alpha-\beta$ (°) | 176–168 |
| Incident angle to detector, $\gamma$ (°) | 140–160 |

As emphasized earlier, the spectroscopic optics of the 2D-RIXS spectrometer are fundamentally identical to those of a conventional spectrometer, with the exception that the dispersion plane is rotated by 90° around the optical axis. Accordingly, the spectrometer types follow the classifications in Ref. [10], and the optimization methods follow the approach outlined in Ref. [20]. Since the optical parameters required to achieve ultrahigh resolution are mathematically determined by these boundary conditions and methodologies, the primary focus of our design effort was on the practical implementation, specifically regarding the mechanical configuration and the maximization of grating performance.

The first major design decision concerned the optical layout: whether to adopt a simple configuration consisting only of a grating and a detector, or a variable included angle configuration by adding a plane mirror upstream of the grating (as seen in the spectrometer of SIX at NSLS-II). From an efficiency standpoint, the variable included angle design is disadvantageous because the additional reflection by the plane mirror reduces throughput. For this reason, a simple layout is preferred. However, in a conventional vertical dispersion spectrometer, a simple design can suffer from mechanical instability because the detector must be elevated significantly to track the diffracted X-rays as the energy changes. In such cases, the variable included angle design, which allows for a constant outgoing beam direction, offers significant stability advantages by lowering the center of gravity and reducing vertical motion. In the case of 2D-RIXS, however, the grating disperses X-rays horizontally. The detector moves on a horizontal plane supported by the floor, eliminating the need to elevate the detector arm to significant heights. Consequently, the stability benefit of the variable included angle design is less compelling for the 2D-RIXS spectrometer. Prioritizing measurement efficiency, we thus adopted the simple configuration consisting only of a diffraction grating and a detector.

To achieve the target ultrahigh resolution ($E/\Delta E>100,000\times\sqrt{2}$), the performance of the diffraction grating, including both the substrate and the groove, is critical. Regarding the substrate shape, cylindrical gratings are commonly used in RIXS spectrometers to prevent the focal plane from becoming too grazing (too parallel to the beam). We initially considered this standard approach. However, we found that in the extreme ultrahigh-resolution regime targeted by the 2D-RIXS spectrometer, the focal plane of even a plane grating becomes sufficiently upright, with an incident angle to the detector surface of approximately 20°–40°. (Note that in our geometry, the detector faces the grating, corresponding to a $\gamma$ of 140°–160° in standard notation). From the perspective of fabrication, plane substrates can be manufactured with higher surface accuracy and quality than cylindrical ones. Therefore, taking advantage of this unique characteristic of the ultrahigh-resolution regime, we adopted a plane substrate for the highest-resolution grating to maximize optical quality, while retaining cylindrical substrates for other resolution modes.

Regarding the grating grooves, achieving $E/\Delta E>100,000\times\sqrt{2}$ with first-order diffraction would require an exceedingly high groove density of approximately 3000 lines/mm. Generally, higher groove density leads to lower diffraction efficiency, potentially making the throughput a limiting factor. Even employing a blazed profile instead of a laminar one is often insufficient to recover the efficiency at such high densities. To address this, we adopted a strategy of using higher-order diffraction, similar to the approach used for the beamline monochromator [27]. Figure 6 shows the comparison of calculated efficiency for three scenarios: 3000 lines/mm (1st order), 1500 lines/mm (2nd order), and 1000 lines/mm (3rd order). The diffraction efficiency was calculated by the differential method [28]. The results indicate that while higher orders result in a narrower high-efficiency energy range, the peak intensity is higher. Given that covering a wide energy range (e.g., 250–2000 eV) with a single grating is mechanically impractical and the target range for one grating is effectively limited to 500–1000 eV, a "narrow but strong" efficiency profile is better matched to the scientific and mechanical requirements than a "wide but weak" one. A lower groove density (1000 lines/mm) lowers the manufacturing difficulty, leading to higher overall quality. Based on these considerations, we selected a groove density of 1000 lines/mm used in the 3rd diffraction order. In the ultrahigh-resolution mode, the source (undulator), beamline monochromator, and spectrometer operate at the 1st, 2nd, and 3rd orders, respectively.

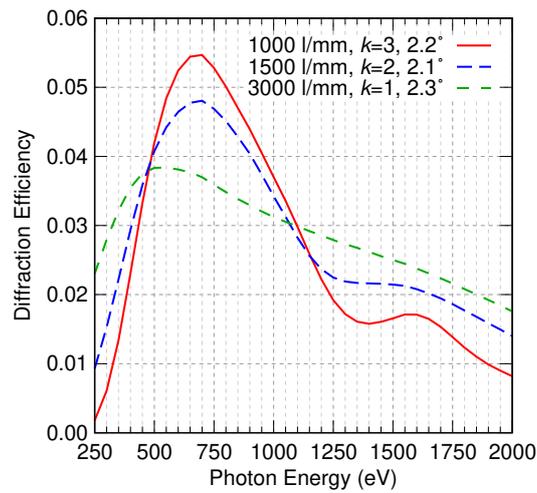

**FIG. 6.** Comparison of calculated diffraction efficiencies for three grating scenarios: 3000 lines/mm (1st order), 1500 lines/mm (2nd order), and 1000 lines/mm (3rd order).

For the high-resolution gratings, plane substrates result in an impractically grazing focal plane; thus, standard cylindrical gratings were employed. The parameters for the gratings are summarized in Table V. Figure 7 plots the resolution and configuration parameters as a function of energy. Calculations confirm that the Spec-G1000 plane grating achieves $E/\Delta E > 100{,}000 \times \sqrt{2}$ below 1000 eV. Notably, the required $r_1$ movement for the Spec-G1000 plane grating is notably smaller than that for the Spec-G1400 cylindrical grating. Although the use of a plane grating is unconventional for RIXS spectrometers, it became feasible here as a consequence of the ultrahigh-resolution conditions. The plane grating proved to be a superior choice, offering advantages not only in minimizing substrate slope error but also in simplifying mechanical requirements. These two designs satisfy the mechanical boundary conditions for $r_1$, included angle ($\alpha-\beta$), and detector angle $\gamma$.

**TABLE V.** Grating parameters of the 2D-RIXS spectrometer. Values in brackets [ ] are the measured data.

|  | Spec-G1000 | Spec-G1400 |
|---|---|---|
| Energy range (eV) | 250–1100 | 250–1000 |
| Reference energy (eV) | 700 | 1000 |
| Incident angle at reference energy (°) | 88.95 | 88.6 |
| Entrance arm length at reference energy (m) | 3.45 | 3.5 |
| Exit arm length at reference energy (m) | 8.55 | 8.5 |
| Focal plane angle at reference energy (°) | 155.1 | 158 |
| Shape of substrate | Plane | Cylinder |
| Meridional radius | >100 km [∞] | 292.057 m |
| Sagittal radius | >20 km [∞] | >20 km [∞] |
| Meridional slope error (μrad in RMS) | <0.05 [0.039] | <0.1 [0.058] |
| Sagittal slope error (μrad in RMS) | <0.5 [0.048] | <0.5 [0.113] |
| Groove profile | Blazed | Laminar |
| Diffraction order | 3 | 1 |
| Ruled area $L \times W$ (mm$^2$) | 180×20 | 180×20 |
| Blaze angle (°) | 2.2 [2.04] | — |
| Duty ratio | — | 0.3 [0.41] |
| Groove depth (nm) | — | 8 [8.6] |
| Material of substrate | Si | Si |
| Material of coating | Au | Au |
| VLS parameters: |  |  |
| $a_0$ (lines/mm) | 1000 | 1400 [1400.120] |
| $a_1$ (lines/mm$^2$) | 0.259 | 0.2796 [0.2793] |
| $a_2$ (lines/mm$^3$) | 3.40×10$^{-5}$ | 6.853×10$^{-6}$ [7.760×10$^{-6}$] |
| $a_3$ (lines/mm$^4$) | 9.57×10$^{-9}$ | 1.91×10$^{-8}$ |
| Manufacturer | JTEC/Inprentus | JTEC/Shimadzu |

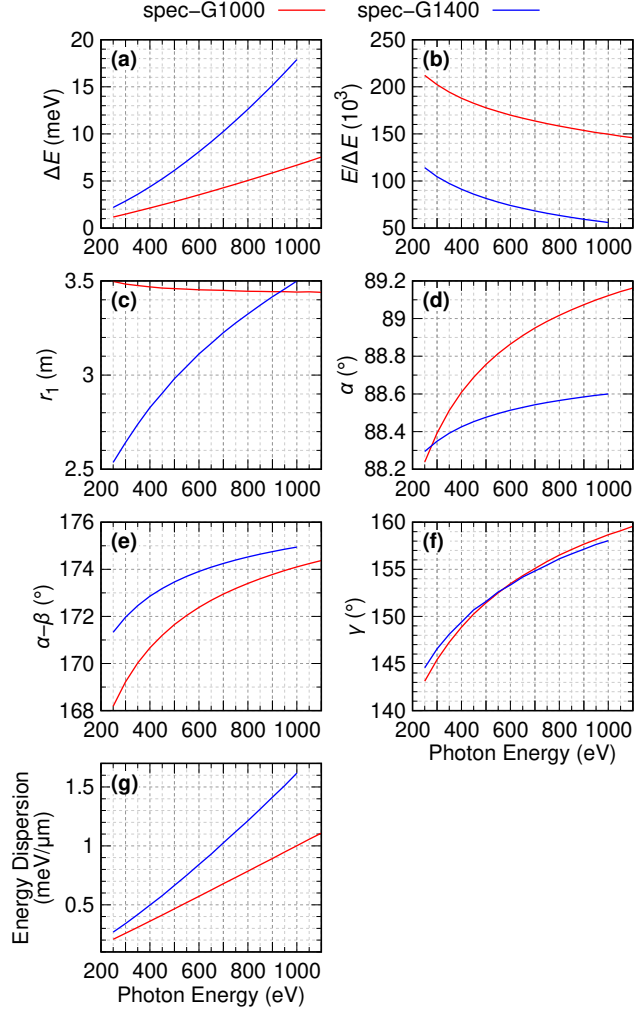

**FIG. 7.** Calculated performance and optical configuration parameters of the spectrometer as a function of photon energy. (a) Energy resolution, $\Delta E$. (b) Resolving power, $E/\Delta E$. (c) Entrance arm length, $r_1$. $r_2$ is determined by the constraint $r_1+r_2=12$ m. (d) Incident angle, $\alpha$. (e) Included angle, $\alpha-\beta$. (f) Incident angle to the detector, $\gamma$. (g) Energy dispersion.

Finally, we examine the focal plane characteristics on the charge-coupled device (CCD) detector, as shown in Fig. 8. At an incident energy of 1000 eV, the detectable energy range covers approximately ±12.5 eV for the ultrahigh-resolution setup and ±20 eV for the high-resolution setup across the full width of the detector. Within a range of ±10(20) eV for the ultrahigh (high) resolution mode, the degradation in resolution is limited to approximately 2(4)%, indicating that the spectrometer maintains consistent performance across this wide energy window. Such a flat focal plane is realized over the 500–1000 eV range (data not shown). As the incident energy decreases, the observable spectral range narrows proportionally. At 500 eV, the full detector width covers ±5.5 eV (8 eV) for the ultrahigh (high) resolution mode. Even at this lower energy, the energy range of ±5.5 eV can be effectively used with acceptable resolution degradation in either mode. Since the 2D-RIXS spectrometer employs spectroscopic optics similar to conventional spectrometers, it offers comparable utility regarding the energy loss range. This ensures that the incident energy dependence of RIXS spectra can be measured over a wide range, allowing for the simultaneous observation of low-energy excitations near the elastic peak as well as higher-energy features like *dd* or charge-transfer excitations.

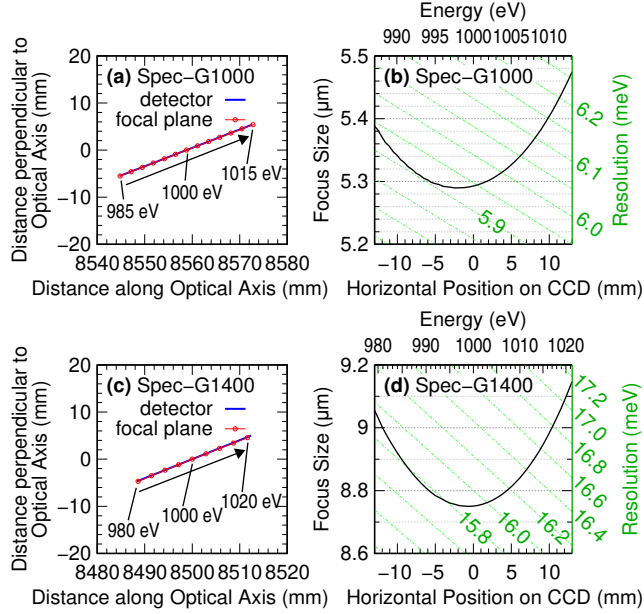

**FIG. 8.** Calculated focal plane characteristics and energy resolution around 1000 eV of the 2D-RIXS spectrometer. (a) and (b) Ray-tracing simulation of the focal plane at the detector for Spec-G1000 and Spec-G1400, respectively. (b) and (d) Horizontal position dependence of the focus size on the detector around 1000 eV for Spec-G1000 and Spec-G1400, respectively. The focus size is also expressed in the energy scale.

### C. Imaging Optics in 2D-RIXS Spectrometer

In the original $h\nu^2$ concept proposed by Strocov, a simple elliptical mirror was suggested as the imaging element. However, the imaging quality of a single elliptical mirror is poor due to severe aberrations, making it unsuitable for practical applications. To achieve high-quality imaging, an optical system that satisfies the Abbe sine condition is required [31]. For this reason, spectrometers at ALS and European XFEL have adopted Wolter type I mirrors [32,15,16]. For the 2D-RIXS spectrometer at BL02U, we conducted a comparative study between standard Wolter mirrors, which satisfy the Abbe sine condition approximately, and Wolter-Schwarzschild (WS) mirrors, which satisfy it exactly [33,34].

Wolter mirrors are classified into three types: I, II, and III. Since the imaging optics of a 2D-RIXS spectrometer functions as a magnifying microscope, the optical design considerations are inverse to those of demagnifying Wolter telescope systems. In this magnifying context, the Type III configuration places the intersection of the elliptical and hyperbolic surfaces (the point determining magnification) closer to the object. This geometry is advantageous for applications requiring high (de)magnification and a long working distance [35]. However, for the 2D-RIXS spectrometer, a long working distance reduces the acceptance angle, thereby degrading the capability as a collection mirror. Hence, Type III is unsuitable for the imaging optics of this spectrometer. Therefore, our investigation focused on Wolter types I and II, as well as the WS design.

#### 1. Wolter Mirror Geometry

We first define the geometry of the Wolter mirror. A Wolter mirror consists of a combination of an ellipsoid and a hyperboloid that are coaxial and confocal. The configurations for Types I and II are illustrated in Fig. 9(a) and (b), respectively. Since we focus on 1D imaging, the problem reduces to a 2D geometry. Accordingly, we will refer to the optical surfaces as a hyperbola and an ellipse in the subsequent discussion. We consider a magnifying imaging setup where X-rays originating from the object $F_o$ are reflected sequentially by the hyperbola and the ellipse, finally focusing at the image point $F_i$. $F_v$ denotes the virtual focus shared by the two surfaces. For the mathematical description, we define the coordinate systems as follows: for the hyperbola, the origin is at $F_o$ with the positive axis pointing towards $F_i$, while for the ellipse, the origin is at $F_i$ with the positive axis pointing towards $F_o$. The polar angles $\theta_3$ (centered at $F_o$) and $\theta_1$ (centered at $F_i$) are defined as positive in the clockwise and counter-clockwise directions, respectively. In Type I, the first reflection occurs on the concave surface of the hyperbola, whereas in Type II, it occurs on the convex surface. Defining the shapes and relative positions of the ellipse and hyperbola requires a total of five parameters (geometrically, four parameters for the semi-axes $a$ and $b$ of the hyperbola and ellipse and one for their relative position). In this study, we selected the following five parameters for design and optimization: the magnification $M$, the working distance $W$ (defined as the distance from $F_o$ to the entrance edge of the mirror), the grazing incident angle at the mirror edge $\theta_{3w}$, the distance between the object and image foci $d$, and the total optical path length $s$.

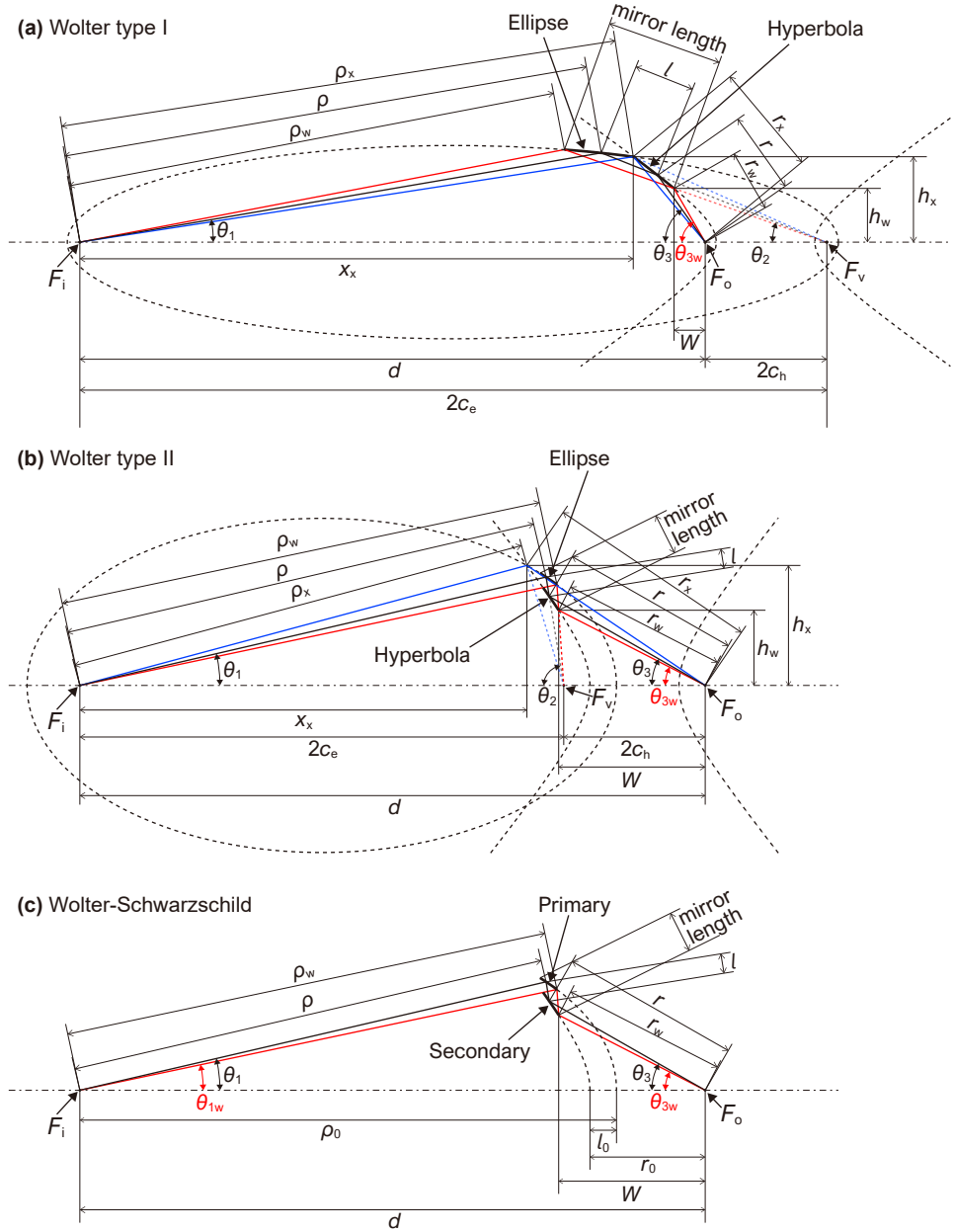

**FIG. 9.** Schematic geometries of the three microscope mirror configurations: (a) Wolter type I, (b) Wolter type II, and (c) Wolter-Schwarzschild.

In the following, we describe the derivation of the equations that define the Wolter mirror geometry using these five parameters. Let $r$, $l$, and $\rho$ be the optical path lengths from $F_o$ to the hyperbola reflection point, from the hyperbola to the ellipse reflection point, and from the ellipse to $F_i$, respectively. The total path length is given by $s=r+l+\rho$. Let $r_w$ be the distance from $F_o$ to the entrance edge of the hyperbola, and let $r_x$ and $\rho_x$ be the distances from $F_o$ to the intersection point of the two surfaces and from the intersection point to $F_i$, respectively.

From the definitions, $r_w$ is given by

$$r_w = \frac{W}{\cos\theta_{3w}}.$$

Since the intersection point lies on the optical path, the relation $r_x + \rho_x = s$ holds. Defining the magnification as $\rho_x = Mr_x$, we obtain

$$r_x = \frac{s}{1+M}, \quad \rho_x = \frac{Ms}{1+M}.$$

The height of the intersection point from the optical axis, $h_x$, and the axial distance from $F_i$ to the intersection point, $x_x$, can be calculated using Heron's formula for the triangle formed by $F_o$, $F_i$, and the intersection point (with side lengths $d$, $r_x$, $\rho_x$)

$$h_x = \frac{2\sqrt{\frac{d+s}{2}\left(\frac{d+s}{2}-d\right)\left(\frac{d+s}{2}-\rho_x\right)\left(\frac{d+s}{2}-r_x\right)}}{d}, \quad x_x = \sqrt{\rho_x^2 - h_x^2}.$$

Using these values, we determine the geometric parameters for the hyperbola (subscript h) and ellipse (subscript e): the semi-major axis $a$, semi-minor axis $b$, distance from center to focus $c$, eccentricity $e$, and distance from focus to directrix $p$. For the hyperbola,

$$e_h = \frac{r_x - r_w}{d - x_x - W}$$

$$p_h = \frac{r_w}{e_h} - W$$

$$a_h = \frac{e_h}{e_h^2 - 1} p_h$$

$$c_h = e_h a_h$$

$$b_h = \sqrt{c_h^2 - a_h^2}.$$

For the ellipse,

$$c_e = \frac{d + 2c_h}{2}$$

$$a_e = \frac{\left(\rho_x + \sqrt{(2c_e - x_x)^2 + h_x^2}\right)}{2}$$

$$b_e = \sqrt{a_e^2 - c_e^2}$$

$$e_e = \frac{c_e}{a_e}$$

$$p_e = \frac{a_e(1 - e_e^2)}{e_e}.$$

Then, for an arbitrary polar angle $\theta_3$ around $F_o$, the corresponding polar angles $\theta_2$ (around the common focus, $F_v$) and $\theta_1$ (around $F_i$) are given by

$$\theta_2 = 2\operatorname{atan}\left(\frac{\frac{e_h - 1}{e_h + 1}}{\tan\left(\frac{\theta_3}{2}\right)}\right), \quad \theta_1 = 2\operatorname{atan}\left(\frac{\frac{1 - e_e}{1 + e_e}}{\tan\left(\frac{\theta_2}{2}\right)}\right).$$

The radial distances $r$ and $\rho$ in polar coordinates centered at $F_o$ and $F_i$, respectively, are expressed as

$$r(\theta_3) = \frac{e_h p_h}{1 - e_h \cos\theta_3}, \quad \rho(\theta_1) = \frac{e_e p_e}{1 - e_e \cos\theta_1}.$$

Finally, the grazing incident angles $i_h$ and $i_e$ at each reflection point are given by:

$$\tan i_h = -\frac{1}{r}\frac{dr}{d\theta_3}, \quad \tan i_e = -\frac{1}{\rho}\frac{dr}{d\theta_1}.$$

It is important to note that both Type I and Type II configurations can be described continuously using the parameters $c$, $a$, $b$, $e$, and $p$ for the hyperbola and ellipse. The boundary value $d_{\text{I-II}}$, which defines the transition where the reflection on the hyperbola switches between concave and convex, is given by

$$d_{\text{I-II}} = r_x \cos\theta_{3w} + \sqrt{\rho_x^2 - (r_x \sin\theta_{3w})^2}.$$

## 2. Wolter -Schwarzschild Mirror Geometry

Although the geometry of the WS mirror for microscopy applications is described in Ref. [36], for the purpose of comparison with the Wolter mirror described above, we redefine the geometry using the same set of five parameters shown in Fig. 9(c): $M$, $W$, $\theta_{3w}$, $d$, and $s$. However, we slightly modify the definition of magnification $M$ for the WS configuration. Let $m_1$ be the magnification defined from the primary to the secondary surface (where $m_1<1$ corresponds to demagnification), and $m_2 = 1/m_1$ be the magnification defined from the secondary to the primary surface. Thus, $m_2$ corresponds to the magnification $M$ used in the previous definitions for the Wolter mirror. The other parameters remain the same: the working distance $W$, the grazing incident angle at the mirror entrance $\theta_{3w}$, the distance between foci $d$, and the total optical path length $s$.

Since the WS mirror strictly satisfies the Abbe sine condition, the relation

$$\sin\theta_{1w} = m_1 \sin\theta_{3w}$$

holds, which allows $\theta_{1W}$ to be determined. $r_w$ is simply given by

$$r_w = \frac{W}{\cos\theta_{3w}}.$$

In this case, $\rho_w$ is expressed as

$$\rho_w = \frac{(d-W)^2 + W^2\tan^2\theta_{3w} - (s-r_w)^2}{2\big((d-W)\cos\theta_{1w} + W\sin\theta_{1w}\tan\theta_{3w} - (s-r_w)\big)}.$$

Here, the initial conditions $\rho_0$, $l_0$, and $r_0$ are shown in Fig. 9(c). Based on this geometry, these parameters are related to our parameters by $l_0 = \frac{s-d}{2}$, $\rho_0 + r_0 = \frac{s+d}{2}$. Thus, $\kappa$ is defined as

$$\kappa = \frac{\rho_0 + r_0}{l_0} = \frac{s+d}{s-d}.$$

By defining the auxiliary parameters as

$$\alpha = \frac{m_1\kappa}{m_1\kappa - 1}, \beta = \frac{m_1}{m_1 - \kappa}, \alpha' = \frac{m_2\kappa}{m_2\kappa - 1}, \beta' = \frac{m_2}{m_2 - \kappa},$$

$$\gamma(\theta_1) = \cos\theta_1 + m_1\sqrt{1 - \left(\frac{\sin\theta_1}{m_1}\right)^2}, \delta(\theta_3) = \cos\theta_3 + m_2\sqrt{1 - \left(\frac{\sin\theta_3}{m_2}\right)^2},$$

the functions for the primary and secondary surfaces, $\rho(\theta_1)$ and $r(\theta_3)$, are given by

$$\frac{l_0}{\rho(\theta_1)} = \frac{1+\kappa}{2\kappa} + \frac{1-\kappa}{2\kappa}\cos\theta_1$$
$$+ C_p\gamma(\theta_1)^{-1}\big(\gamma(\theta_1) - (1-m_1)\big)^{\alpha}\big(\gamma(\theta_1) - (m_1-1)\big)^{\beta}\big((\kappa+1)\gamma(\theta_1) - (\kappa-1)(m_1+1)\big)^{2-\alpha-\beta},$$

$$\frac{l_0}{r(\theta_3)} = \frac{1+\kappa}{2\kappa} + \frac{1-\kappa}{2\kappa}\cos\theta_3$$
$$+ C_s\delta(\theta_3)^{-1}\big(\delta(\theta_3) - (1-m_2)\big)^{\alpha'}\big(\delta(\theta_3) - (m_2-1)\big)^{\beta'}\big((\kappa+1)\delta(\theta_3) - (\kappa-1)(m_2+1)\big)^{2-\alpha'-\beta'}.$$

Here, $C_p$ and $C_s$ are integration constants determined from the boundary values $\rho_w$ and $r_w$ at $\theta_{1w}$ and $\theta_{3w}$, respectively:

$$C_p = \frac{\left(\frac{l_0}{\rho_w} - \left(\frac{1+\kappa}{2\kappa} + \frac{1-\kappa}{2\kappa}\cos\theta_{1w}\right)\right)}{\gamma(\theta_{1w})^{-1}\big(\gamma(\theta_{1w}) - (1-m_1)\big)^{\alpha}\big(\gamma(\theta_{1w}) - (m_1-1)\big)^{\beta}\big((\kappa+1)\gamma(\theta_{1w}) - (\kappa-1)(m_1+1)\big)^{2-\alpha-\beta}},$$

$$C_s = \frac{\left(\frac{l_0}{r_w} - \left(\frac{1+\kappa}{2\kappa} + \frac{1-\kappa}{2\kappa}\cos\theta_{3w}\right)\right)}{\delta(\theta_{3w})^{-1}\big(\delta(\theta_{3w}) - (1-m_2)\big)^{\alpha'}\big(\delta(\theta_{3w}) - (m_2-1)\big)^{\beta'}\big((\kappa+1)\delta(\theta_{3w}) - (\kappa-1)(m_2+1)\big)^{2-\alpha'-\beta'}}.$$

## 3. Optimization of Mirror Geometry

Based on these formulations, we optimized the mirror geometry for the 2D-RIXS spectrometer. The total path length was fixed at $s=12$ m. To maximize the acceptance angle, the working distance was set to $W=100$ mm, placing the mirror as close to the sample as possible. A maximum mirror length of 400 mm was imposed to fit within the vacuum chamber. The magnification $M$ is largely determined by this geometry; we set $M=43$ for the Wolter type I and $M=22$ for the Wolter type II and WS. We evaluated the designs based on two criteria: (1) the ability to collect scattered X-rays (effective acceptance angle, defined as the product of physical acceptance and reflectivity) and (2) the ability to image over a wide range with a spatial resolution of ≤2 μm (effective FOV). The efficiency of the 2D-RIXS spectrometer depends on both, so we defined a Figure of Merit (FOM) as their product.

The variable parameters were $d$ and $\theta_{3w}$, which effectively determine the curvatures and incident angles of the two mirrors. Figure 10 shows the dependence of the effective acceptance angle, effective FOV, and FOM on these two parameters at 900 eV. Notably, the trends for the Wolter type II and WS are very similar, confirming that Wolter mirrors satisfy the Abbe sine condition well in grazing incidence geometries [37].

The effective acceptance angle depends on the trade-off between the physical acceptance angle (which increases as $\theta_{3w}$ increases and the mirror becomes more "upright") and reflectivity (which decreases as the grazing angle increases). For type I, the dependence of the effective acceptance angle on $d$ for each $\theta_{3w}$ exhibits a sharp peak (Fig. 10(a)), because the 400 mm mirror length limits the geometrical acceptance; the peak corresponds to the configuration where the lengths of the hyperbolic and elliptical surfaces are optimally balanced, and thus both surfaces are fully utilized. The effective acceptance peaks around $\theta_{3w}\approx7°$ at 900 eV. In contrast, for type II/WS, the limiting factor is not the mirror length itself but the geometric constraint on the hyperbola length, which must be restricted to avoid blocking the reflection from the ellipse. This leads to a smoother dependence peaked around $\theta_{3w}\approx1.2°$ (Fig. 10(d) and (g)). Generally, type I offers a larger acceptance angle than type II/WS.

Regarding the effective FOV, type I shows a monotonic increase with $\theta_{3w}$ (Fig. 10(b)), which is attributed to the relaxation of grazing incidence conditions reducing aberrations. Conversely, type II/WS exhibits a distinct peak (Fig. 10(e) and (h)), indicating an optimal configuration where astigmatism and field curvature from the convex and concave surfaces cancel out. Since the WS design is inherently free of coma aberration, this peaked behavior suggests that the higher-order aberrations are minimized at the specific geometry.

The resulting FOM maximums of type II/WS are better than that of type I. At 900 eV, the optimum is $\theta_{3w}\approx10°$ for type I (Fig. 10(c)) and $\approx0.5°$ for type II/WS (Fig. 10(f) and (g)). Figure 10(j) and (k) also present the results for type I at 250 eV and 2000 eV, which correspond to the lower and upper limits of the energy range. At 250 eV, since high reflectivity is maintained even at larger grazing angles, the FOM continues to increase as $\theta_{3w}$ increases from 3° to 15°. Conversely, at 2000 eV, since reflectivity drops sharply as the angle deviates from the grazing condition, the FOM decreases rapidly at large $\theta_{3w}$, peaking at $\theta_{3w}\approx6°$. For type II/WS, the optimum remains around 0.5° even at 2000 eV (not shown here). Considering the energy range up to 2000 eV, the optimal angles are approximately 6° for type I and 0.5° for type II/WS. From a construction standpoint, a large $\theta_{3w}$ in the Type I geometry (downward reflection) pushes the detector position significantly downwards. For a 12 m arm, $\theta_{3w}>5°$ would require the detector to be below the floor level, making construction difficult. Therefore, $\theta_{3w}\approx5°$ becomes a practical limit.

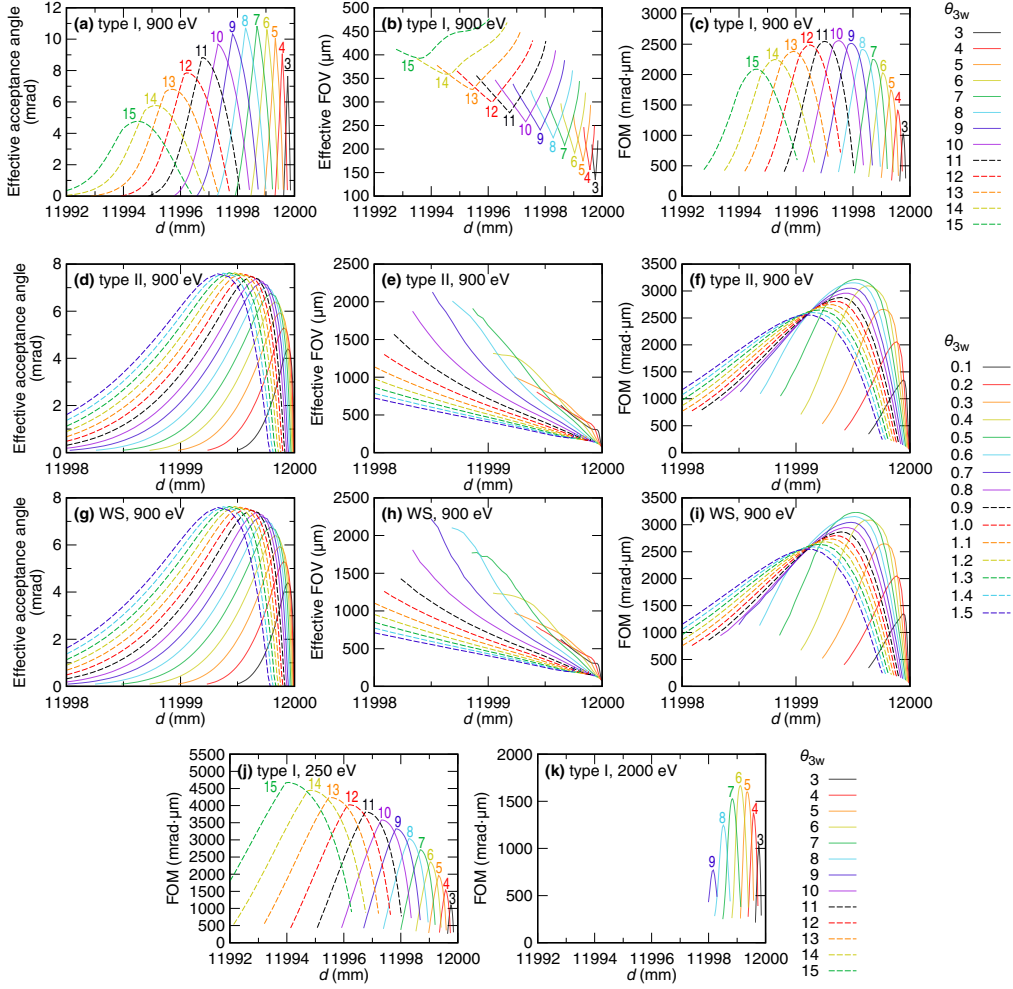

**FIG. 10.** Dependence of the effective acceptance angle, effective FOV, and FOM on the geometric parameters $d$ and $\theta_{3w}$. (a)–(c) Results for Wolter type I at 900 eV. (d)–(f) Results for Wolter type II at 900 eV. (g)–(i) Results for Wolter-Schwarzschild at 900 eV. (j), (k) FOM dependence for Wolter type I at 250 eV and 2000 eV, respectively. The ranges of $\theta_{3w}$ are 3°–15° in 1° steps for Wolter type I and 0.1°–1.5° in 0.1° steps for Wolter type II/WS.

Ultimately, we selected the monolithic Wolter Type I design ($M=43$, $W=100$ mm, $\theta_{3w}=5°$). Although type II/WS offers a larger effective FOV, type I was chosen for its (1) ease of alignment and stability due to the monolithic fabrication, and (2) better matching with the 2D-RIXS strategy. A larger effective FOV implies collecting scattering from a wider incident energy range, which could degrade the effective beamline resolution (due to the beamline focal plane) and introduce non-negligible incident energy dependence, complicating data integration. Furthermore, even assuming the incident energy dependence is acceptable, using a wider effective FOV demands sample homogeneity over a larger area, thereby reducing the flexibility of experiments. The characteristics of the Wolter type I, specifically its high effective acceptance with a moderate effective FOV, are more suitable for the 2D-RIXS spectrometer.

The final parameters are $s=12000$ mm and $d\approx 11999.3$ mm. The mirror parameters are summarized in Table VI. Figure 11 shows the imaging performance (spatial resolution dependence on the sample position) for these parameters. The Wolter mirror achieves a spatial resolution of <2 µm over a range of ±80 µm, satisfying the ultrahigh-resolution requirements. With the detector tilted at 20–40°, the optical path length varies by approximately ±11.7 mm across the 25 mm detector surface. This variation could induce a degradation in imaging performance due to defocusing along the optical axis. However, as shown in Fig. 11, ray-tracing simulations confirm that the degradation in spatial resolution is negligible even with a ±12 mm defocus. The resolution remains <2 µm over the effective FOV of ±80 µm at any position on the tilted detector.

TABLE VI. Mirror parameters of the 2D-RIXS spectrometer. Values in brackets [ ] are the measured data.

| Shape | Wolter type I | |
|---|---|---|
| | Hyperbola | Ellipse |
| Entrance arm length (mm) | 150.72 | 520.85 |
| Exit arm length (mm) | 290.43 | 11618.86 |
| Angle of incidence (°) | 88.92 | 88.78 |
| Meridional radius | a=69.856, b=3.944 | a=6069.86, b=52.191 |
| Sagittal radius | — | — |
| Material of substrate | Si | |
| Meridional slope error (μrad in RMS) | <0.15 [0.115] | <0.15 [0.106] |
| Sagittal slope error (μrad in RMS) | <1.0 [0.298] | <1.0 [0.125] |
| Material of coating | Au | Au |
| Useful area $L \times W$ (mm$^2$) | ~217×20 | ~161×20 |
| Manufacturer | JTEC | |

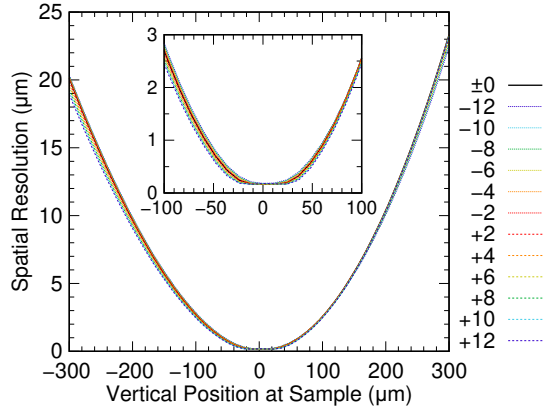

FIG. 11. Spatial resolution dependence for the optimized Wolter Type I mirror, M4. The plot shows the spatial resolution as a function of the position at the sample. The inset shows a magnified plot ranging from −100 μm to +100 μm. The black thick line (±0) represents the performance at the best focus position, while the colored dotted and dashed lines indicate the performance with a defocus of −12 mm to −2 mm (2 mm step) and +2 mm to +12 mm (2 mm step), corresponding to the optical path length variation across the tilted detector surface. The spatial resolutions remain below 2 μm over the effective FOV of ±80 μm.

## D.  Expected Total Performance

First, we calculated the theoretical total energy resolution and resolving power by combining the beamline performance (Sec. IV A) and the spectrometer performance (Fig. 7). The total resolution is defined as the root-sum-square of the beamline and spectrometer contributions. The results are plotted in Fig. 12(a) and (b). The calculation confirms that the design achieves a resolving power over 100,000 in the target energy range of 500–1000 eV, demonstrating the ultrahigh resolution capability of the 2D-RIXS facility.

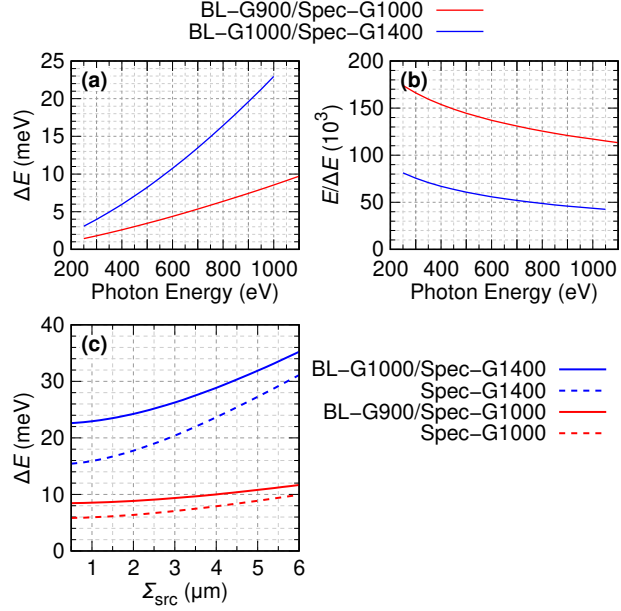

**FIG. 12.** Calculated total (a) energy resolution and (b) resolving power of the 2D-RIXS spectrometer. The red lines represent the ultrahigh-resolution mode (BL-G900 and Spec-G1000), and the blue lines represent the high-resolution mode (BL-G1000 and Spec-G1400).

The calculated total energy resolution at 1000 eV is approximately 8.5 meV for the ultrahigh-resolution mode. These values might appear to have a large margin compared to the <10 meV target. This margin is intentional, as the simulation assumed a fixed source size of $\Sigma_{src}$=1 μm. In our 2D-RIXS configuration, the effective source size is determined by the horizontal width of the beam footprint on the sample, which varies with the incident angle during experiments. Figure 12(c) shows the dependence of the spectrometer and total energy resolutions on $\Sigma_{src}$ at 1000 eV. Although a larger $\Sigma_{src}$ degrades the spectrometer resolution, the total resolution is maintained <10 meV for source sizes up to approximately 4 μm in the ultrahigh-resolution mode. This means that the total energy resolution can be maintained <10 meV even at a grazing incidence of θ≈10°, where the focused beam size of about 0.7 μm expands to about 4.0 μm on the sample.

We then evaluated and verified the total energy resolution of the 2D-RIXS spectrometer at 1000 eV using full ray-tracing simulations (SHADOW) based on the determined parameters for the Wolter Type I mirror (M4) and the spectrometer gratings. The effective source for the spectrometer ray-tracing was derived directly from the beamline ray-tracing results. Specifically, the focused beam profile and the spatial distribution of the dispersed energy on the sample were converted into the source definition for the spectrometer ray-tracing. The horizontal focus size was set at 1 μm. The scattering from the sample was assumed to be isotropic with uniform divergence. Regarding optical imperfections, slope errors of 0.15 μrad for the Wolter mirror (M4) and 0.05 μrad for the grating were included, reflecting their respective design specifications.

Figure 13(a) displays the resulting footprint of the elastic scattering on the detector (CCD) obtained by the ray-tracing for the ultrahigh resolution mode using BL-G900 and Spec-G1000. This footprint represents the purely geometrical distribution of where the energy-dispersed X-rays are focused and imaged. At the sample, the incident X-rays are vertically energy dispersed, meaning the incident photon energy varies with the vertical position. The Wolter mirror images this vertical profile onto the detector, mapping the incident energy distribution to the vertical direction of the detector. Simultaneously, the spectrometer grating disperses the X-rays horizontally according to their energy. Because the energy varies with the vertical position, the horizontal focusing position determined by the dispersion of the spectrometer grating shifts as a function of the vertical position. As a result, the elastic scattering forms a diagonal line on the detector.

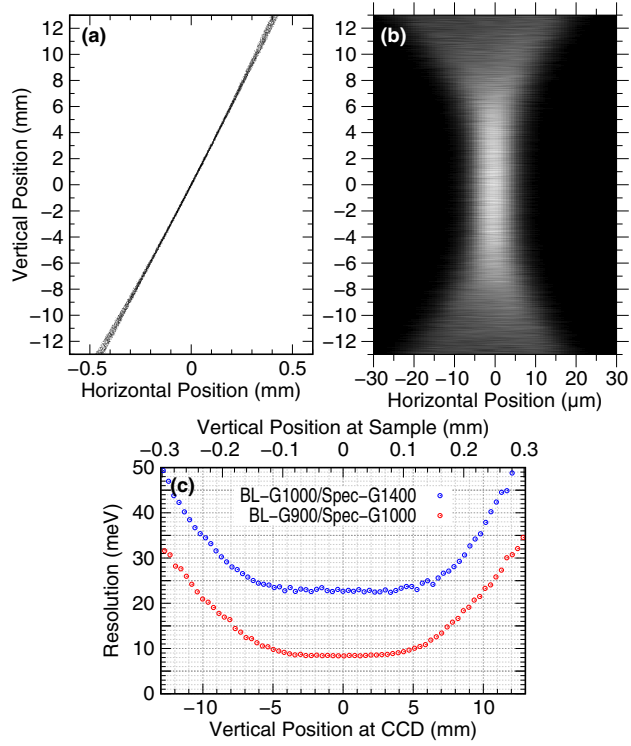

**FIG. 13.** Ray-tracing results for the 2D-RIXS spectrometer at 1000 eV. (a), (b) Results for the ultrahigh-resolution mode (using BL-G900 and Spec-G1000). (a) Beam footprint on the detector. (b) Analyzed intensity map plotted as incident energy information (vertical) versus energy loss information (horizontal). The map includes all the broadening contributions from the beamline and the 2D-RIXS spectrometer. (c) Total energy resolution obtained from the fitting at each vertical position. The red and blue circles correspond to the ultrahigh-resolution (BL-G900 and Spec-G1000) and high-resolution (BL-G1000 and Spec-G1400) modes, respectively.

In the actual measurement, the detected signal is further broadened by the spatial resolution of the detector. Accordingly, we convolved the ray-tracing results with a detector spatial resolution of 4 μm (FWHM) and converted the horizontal axis from scattered photon energy to energy loss by applying a horizontal shift dependent on the vertical position. The resulting image, shown in Fig. 13(b), displays the intensity map (vertical: incident energy information vs. horizontal: energy loss information), including all the contributions from the beamline and the 2D-RIXS spectrometer.

The horizontal width of the signal in this map corresponds to the total energy resolution. As evident from the figure, the resolution is sharp near the center but broadens towards the edges. This degradation is primarily due to aberrations in the Wolter mirror (M4) imaging property (Fig. 11). Figure 13(c) shows the FWHM of the peak derived from fitting at each vertical position. For the ultrahigh-resolution setup (BL-G900 and Spec-G1000), a total resolution of <10 meV is achieved near the center. The results for the high-resolution setup (BL-G1000 and Spec-G1400) are also shown in Fig. 13(c), exhibiting a resolution of approximately 23 meV near the center. Both the ultrahigh-resolution and high-resolution results show excellent agreement with the theoretical calculations presented in Fig. 12.

Although the effective FOV of M4, where the spatial resolution is <2 μm, is limited to approximately ±80 μm (Fig. 11), when all the factors are included, other broadening factors also contribute to the total resolution. Consequently, the simulation at 1000 eV indicates that a vertical range of approximately ±100 μm, corresponding to an incident energy bandwidth of about ±0.14 eV (±0.20 eV) for the ultrahigh (high) resolution mode, can be effectively used without noticeable resolution degradation. This demonstrates that the 2D-RIXS setup offers a substantial efficiency gain compared to using monochromatic X-rays. Since the performance comparison in Sec. III C assumed a conservative usable range of ±80 μm, using the ±100 μm or a wider range would further enhance the experimental efficiency.

Finally, we consider the available energy loss range provided by a single setup, which is equally critical for actual experiments. For instance, when investigating the incident energy dependence of RIXS spectra, a narrow energy loss range would severely limit the experimental utility, making the instrument impractical even if the throughput at a specific energy is high. Although the evaluation of the spectroscopic optics alone in Sec. IV B suggested that an energy loss range of ±10(20) eV could be covered without significant resolution degradation for the ultrahigh (high) resolution mode, to rigorously validate this performance, we conducted full ray-tracing simulations.

Figure 14(a) presents the simulated footprints on the detector in the ultrahigh-resolution mode for elastic scattering (incident energy center of 1000 eV) alongside inelastic scattering corresponding to energy losses of ±10 eV. The results clearly demonstrate that even at these energy losses, the scattered X-rays are effectively dispersed, focused, and imaged without noticeable distortion. To evaluate the spectral quality, we applied the horizontal shift and a 4 μm detector spatial resolution blur and integrated the signal over a vertical range of ±100 μm on the sample (corresponding to approximately ±4.3 mm on the detector). The resulting spectra for the three energy positions are shown in Fig. 14(b). Each peak retains a highly symmetric Gaussian profile, confirming the capability of the instrument as a high-performance spectrometer across a wide energy range.

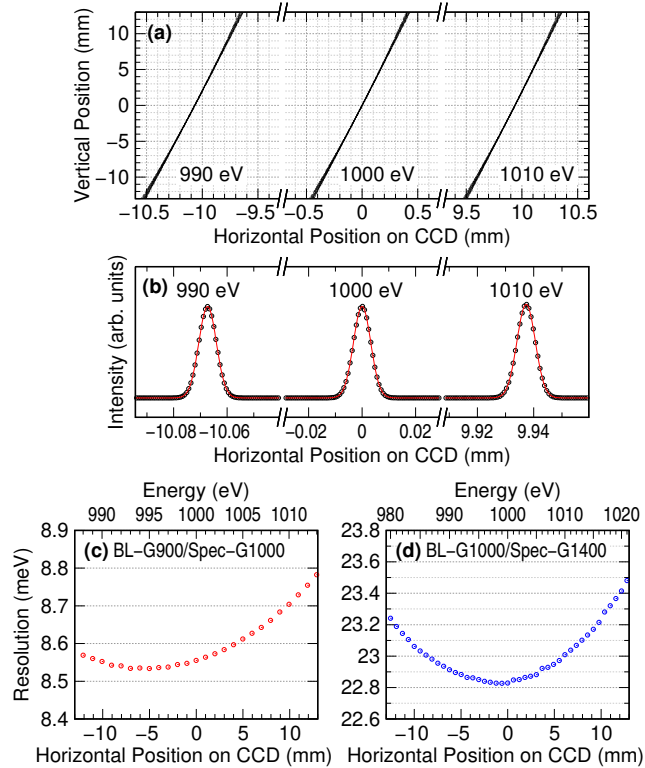

**FIG. 14.** Evaluation of the available energy loss range at 1000 eV. (a) and (b) Results for the ultrahigh-resolution mode. (a) Beam footprints on the detector for elastic scattering (center) and inelastic scattering with ±10 eV energy loss (left and right). Note that the center footprint is identical to that shown in Fig. 13(a). (b) Simulated spectra integrated over a vertical range of ±100 μm on the sample for the three energy positions. (c) and (d) Total energy resolution as a function of the horizontal detector position for the ultrahigh-resolution and high-resolution modes, respectively.

Regarding the vertical integration range, our previous analysis in Sec. IV C set a conservative criterion of ±80 μm that maintains the ultrahigh energy resolution. However, the current simulation employs a wider integration range of ±100 μm. Notably, extending the range from ±80 μm to ±100 μm results in a resolution degradation of only about 1%, which is negligible for practical applications. Even with an integration range of ±140 μm, the degradation remains limited to approximately 5%. This flexibility allows users to optimize the balance between energy resolution and signal intensity according to their specific scientific requirements.

Furthermore, we evaluated the dependence of the total energy resolution on the horizontal detector position across the full width. Figure 14(c) plots the total energy resolution as a function of the horizontal detector position (corresponding to inelastic scattering energy) for the ultrahigh-resolution mode. For the 25 mm square CCD detector used in the 2D-RIXS spectrometer, the resolution degradation remains within approximately 2% even at the edges, confirming that the entire detector area can be effectively utilized. Similar results were obtained for the high-resolution mode, as shown in Fig. 14(d), where the resolution degradation remains within approximately 4% across the entire range, giving an energy loss range of ±20 eV.

As stated at the beginning of Sec. IV, the imaging and spectroscopic optics in the 2D-RIXS spectrometer are orthogonal, allowing them to be designed and evaluated independently. These full ray-tracing results confirm that the total performance is consistent with the predictions derived from the independent optimizations. The successful implementation of the $hv^2$ concept thus achieves ultrahigh energy resolution and high measurement efficiency while maintaining an energy loss range comparable to conventional spectrometers. This combination provides the 2D-RIXS spectrometer with high usability and flexibility for actual experiments.

## E. Required Mechanical Stability

The mechanical stability tolerances required to achieve the target ultrahigh resolution are summarized in Table VII. The stability of the effective source, defined by the scattering point on the sample, is determined by the convolution of the beam pointing stability of the beamline and the physical position stability of the sample itself. Specifically, the horizontal tolerance ($\Delta X$) is largely determined by sample vibration perpendicular to the optical axis, while the longitudinal tolerance ($\Delta Y$) is determined by vibration parallel to the axis. The vertical source position tolerance ($\Delta Z$) is governed by the energy stability of the dispersed beam from the beamline. This translates to a beamline monochromator pitch stability of approximately 20 nrad, imposing extremely stringent stability requirements on the beamline optics. Conversely, maintaining the physical sample position within the focused beam size (0.66 μm) is considered feasible under standard temperature-controlled conditions.

Regarding the spectrometer optics, the vertical position tolerance ($\Delta Z$) for the Wolter mirror (M4) is particularly challenging. Achieving this stability is non-trivial given that the mirror assembly is mounted on a rotatable $2\theta$ stage and includes complex alignment mechanisms. The pitch stability requirement for the spectrometer grating ($\Delta\theta X$) is set at 28 nrad. Achieving this level of angular stability with the similarly strict requirements for the beamline is one of the most significant engineering challenges of this development. For the detector, the grazing incidence geometry amplifies the sensitivity to vertical displacements ($\Delta Z$), necessitating a highly rigid mechanical design for the detector stage.

Crucially, these tolerances must be maintained not only against high-frequency vibrations but also against low-frequency thermal drifts over typical data acquisition periods, which range from 30 minutes to 1 hour. To suppress thermal drift to an acceptable level, the ambient temperature stability must be controlled to approximately 0.01°C. Therefore, establishing a strictly controlled environment, integrating both robust vibration isolation and precise temperature control, is essential for the successful operation of the 2D-RIXS facility.

Although a detailed quantitative evaluation of specific vibration amplitudes and thermal drifts is challenging at this stage, the 2D-RIXS facility has been successfully commissioned and is currently operational. The fact that ultrahigh-resolution RIXS spectra can be routinely obtained indicates that the stringent stability requirements discussed above have been cleared to a level sufficient for high-resolution measurements.

TABLE VII. Mechanical stability tolerances required to achieve ultrahigh resolution with the Spec-G1000 at 1000 eV. Unmarked values indicate an energy shift of 10% of the resolution, while values marked with an asterisk (*) indicate a resolution degradation of 10%. For the source/sample coordinates, the Y-axis is defined along the optical axis, and the X- and Z-axes correspond to the horizontal and vertical directions, respectively. For the optical elements, the coordinate system is defined relative to the optical surface: the X- and Y-axes lie within the surface plane, orthogonal and parallel to the X-ray incidence direction, respectively, and the Z-axis is surface normal. $\Delta X$, $\Delta Y$, and $\Delta Z$ represent translations, while $\Delta\theta X$, $\Delta\theta Y$, and $\Delta\theta Z$ represent rotations around the respective axes (corresponding to pitch, roll, and yaw).

|  | $\Delta X$ | $\Delta Y$ | $\Delta Z$ | $\Delta\theta X$ | $\Delta\theta Y$ | $\Delta\theta Z$ |
|---|---|---|---|---|---|---|
| Source/Sample | 0.66 μm | 100 μm* | 0.54 μm | — | 1.0°* | — |
| Wolter mirror (M4) | — | 28 μm | 0.53 μm | 1.8 μrad | 35 μrad | 0.07° |
| Grating | — | 3.0 μm | 4.9 μm | 28 nrad | 32 μrad | 0.051° |
| CCD | 28 μm | 0.75 μm | 0.30 μm | 0.3°* | 0.06°* | 0.02°* |

## V. Summary

We have reported the optical design and optimization of the ultrahigh-resolution 2D-RIXS spectrometer developed at NanoTerasu BL02U. Our analysis showed that improving energy resolution in a conventional spectrometer using monochromatic X-rays causes a severe reduction in measurement efficiency, limiting practical RIXS measurements at ultrahigh resolution. To address this, we adopted the $h\nu^2$ concept using dispersive incident X-rays after a comparative study with the AGM–AGS. The 2D-RIXS spectrometer effectively decouples the energy resolution from the incident bandwidth, providing an estimated efficiency gain of more than one order of magnitude compared to conventional spectrometers. The realization of this performance relied on two main technical implementations: the use of higher-order diffraction gratings to enhance diffraction efficiency for ultrahigh resolution, and the optimization of Wolter mirrors for high-quality 1D imaging. We also identified mechanical stability against vibration and thermal drift as an essential requirement for achieving the target resolution. The spectrometer has been constructed and successfully commissioned, confirming the validity of the design strategy. The achieved performance demonstrates that the 2D-RIXS spectrometer can reach the <10 meV resolution range while retaining sufficient throughput for practical experiments. This capability will facilitate the study of low-energy excitations and detailed electronic structures in quantum materials.

## Acknowledgements

The author thanks K. Yamamoto and N. Kurahashi for their contributions to the commissioning and upgrading of the beamline and spectrometer. The RIXS experiments were performed using the 2D-RIXS spectrometer at NanoTerasu BL02U with the approval of the Japan Synchrotron Radiation Research Institute (JASRI) (Proposal Nos. 2025A9059 and 2025B9042).

## Conflict of Interest

The authors have no conflicts to disclose.

## Data Availability

The data that support the findings of this study are available from the corresponding author upon reasonable request.

## References

[1] A. Kotani and S. Shin, Rev. Mod. Phys. **73**, 203 (2001).
[2] L. J. P. Ament *et al.*, Rev. Mod. Phys. **83**, 705 (2011).
[3] F. M. F. de Groot *et al.*, Nat. Rev. Methods Primers **4**, 45 (2024).
[4] M. Mitrano *et al.*, Phys. Rev. X **14**, 040501 (2024).
[5] V. N. Strocov *et al.*, J. Synchrotron Rad. **17**, 631 (2010).
[6] Y. Harada *et al.*, Rev. Sci. Instrum. **83**, 013116 (2012).
[7] N. Brookes *et al.*, Nucl. Instrum. Methods A **903**, 175 (2018).
[8] K. J. Zhou *et al.*, J. Synchrotron Rad. **29**, 563 (2022).
[9] A. Singh *et al.*, J. Synchrotron Rad. **28**, 977 (2021).
[10] J. Dvorak *et al.*, Rev. Sci. Instrum. **87**, 115109 (2016).
[11] J. H. Hubbell *et al.*, J. Phys. Chem. Ref. Data **23**, 339 (1994).
[12] H. S. Fung *et al.*, AIP Conf. Proc. **705**, 655 (2004).
[13] V. N. Strocov, J. Synchrotron Rad. **17**, 103 (2010).
[14] C. H. Lai *et al.*, J. Synchrotron Rad. **21**, 325 (2014).
[15] Y. D. Chuang *et al.*, J. Synchrotron Rad. **27**, 695 (2020).
[16] M. Agåker *et al.*, J. Synchrotron Rad. **31**, 1264 (2024).
[17] R. Wang *et al.*, J. Synchrotron Rad. **32**, 1235 (2025).
[18] K. Yamamoto *et al.*, J. Phys.: Conf. Ser. **3010**, 012115 (2025).
[19] J. Miyawaki *et al.*, Synchrotron Radiat. News **38**, 4 (2025).
[20] V. N. Strocov *et al.*, J. Synchrotron Rad. **18**, 134 (2011).
[21] T. Namioka *et al.*, in JSPE Proceedings on Soft X-ray Optics (Japan Society for Precision Engineering, Tokyo, 1997), pp. 185–198.

[22] G. Ghiringhelli *et al.*, Rev. Sci. Instrum. **77**, 113108 (2006).
[23] A. Amorese *et al.*, Nucl. Instrum. Methods A **513**, 322 (2003).
[24] T. Tokushima *et al.*, Rev. Sci. Instrum. **82**, 073108 (2011).
[25] M. Sanchez Del Rio *et al.*, J. Synchrotron Rad. **18**, 708 (2011).
[26] K. J. Zhou *et al.*, J. Synchrotron Rad. **27**, 1235 (2020).
[27] J. Miyawaki *et al.*, J. Phys.: Conf. Ser. **2380**, 012030 (2022).
[28] M. Boots *et al.*, J. Synchrotron Rad. **20**, 272 (2013).
[29] K. Yamamoto *et al.*, arXiv:2511.17966.
[30] R. Reininger, Nucl. Instrum. Methods A **649**, 139 (2011).
[31] E. Abbe, Sitz. Jena. Ges. Med. Naturwiss. **13**, 129 (1879).
[32] T. Warwick *et al.*, J. Synchrotron Rad. **21**, 736 (2014).
[33] H. Wolter, Ann. Phys. **10**, 94 (1952).
[34] H. Wolter, Ann. Phys. **10**, 286 (1952).
[35] J. Yamada *et al.*, Optics Express **27**, 3429 (2019).
[36] A. K. Head, Proc. Phys. Soc. B **70**, 945 (1957).
[37] R. C. Chase and L. P. Van Speybroeck, Appl. Opt. **12**, 1042 (1973).